\documentclass[12pt]{article}%
\usepackage{amsmath}
\usepackage{amsfonts}
\usepackage{amssymb}
\usepackage{graphicx}%
\newcommand{\gd}{\delta}

\newcommand{\gL}{\Lambda}

\newcommand{\gO}{\Omega}

\renewcommand{\hbar}{\overline{h}}

\newcommand{\zbar}{\overline{z}}

\newcommand{\mubar}{\overline{\mu}}

\newcommand{\Omegabar}{\overline{\Omega}}

\newcommand{\Ltilde}{\tilde{L}}

\newcommand{\bC}{\mathbb C}

\renewcommand{\Im}{\mbox{Im~}}

\renewcommand{\d}{\partial}
\newcommand{\half}{\frac{1}{2}}

\newcommand{\ret}{\nonumber \\}

\input epsf.tex
\newcommand{\be}{\begin{equation}}
\newcommand{\ee}{\end{equation}}
\newcommand{\bea}{\begin{eqnarray}}
\newcommand{\eea}{\end{eqnarray}}

\def\zbar{\bar{z}}

\newcommand{\bZ}{{\mathbf{Z}}}

\begin{document}
\bigskip\begin{titlepage}
\begin{flushright}
UUITP-22/04\\
hep-th/0410141
\end{flushright}
\vspace{1cm}
\begin{center}
{\Large\bf Matrix models, 4D black holes and topological strings on non-compact Calabi-Yau manifolds\\}
\end{center}
\vspace{3mm}
\begin{center}
{\large
Ulf   H.\   Danielsson{$^1$},   Martin  E.
Olsson{$^2$},  and Marcel Vonk{$^3$}} \\
\vspace{5mm}
Institutionen f\"or Teoretisk Fysik, Box 803, SE-751 08
Uppsala, Sweden \\
\vspace{5mm}
{\tt
{$^1$}ulf.danielsson@teorfys.uu.se\\
{$^2$}martin.olsson@teorfys.uu.se\\
{$^3$}marcel.vonk@teorfys.uu.se\\
}
\end{center}
\vspace{5mm}
\begin{center}
{\large \bf Abstract}
\end{center}
\noindent
We study the relation between $c=1$ matrix models at self-dual radii and topological strings on non-compact Calabi-Yau manifolds. Particularly the special case of the deformed matrix model is investigated in detail. Using recent results on the equivalence of the partition function of topological strings and that of four dimensional BPS black holes, we are able to calculate the entropy of the black holes, using matrix models. In particular, we show how to deal with the divergences that arise as a result of the non-compactness of the Calabi-Yau. The main result is that the entropy of the black hole at zero temperature coincides with the canonical free energy of the matrix model, up to a proportionality constant given by the self-dual temperature of the matrix model.

\vfill
\begin{flushleft}
October 2004
\end{flushleft}
\end{titlepage}\newpage

\section{Introduction}

\bigskip

Matrix models of the $c=1$ type have mainly found their applications in the study of two-dimensional string theory. In \cite{Takayanagi:2003sm,Douglas:2003up} it was realized that two specific non-perturbatively well defined matrix models actually correspond to type 0B and type 0A string theory. Type 0B is described by the usual bosonic matrix model, with a potential of the form $\sim -x^{2}$, with both sides of the potential filled, while 0A is described by the so called deformed matrix model where the deformation corresponds to adding a term $\sim 1/x^2$ to the inverted oscillator potential. This deformation effectively removes one side of the original potential, thus naturally stabilizing the system non-perturbatively.

It has, however, also been known for a long time that the bosonic $c=1$ matrix model, at the self-dual radius, arises in a completely different context, namely in relation to topological string theory on the deformed conifold \cite{Ghoshal:1995wm}. In fact, it is now known that there are very general relations between topological string theories and matrix models, see for example \cite{Aganagic:2003qj}. In general, one can identify the partition function of a certain matrix model with the partition function of a corresponding topological string theory.

In this context, it is of course intriguing to see what the roles played by the type 0 theories will be. What makes this particularly interesting at the moment are the results of \cite{Ooguri:2004zv}, where it was conjectured that the topological partition function is equivalent to the partition function of four dimensional BPS black holes, according to
\begin{equation} \label{eq:BHtop}
Z_{BH}=|Z_{top}|^2,
\end{equation} 
where the four-dimensional black hole space-time arises after compactifying string theory on a certain Calabi-Yau manifold, and it is this same Calabi-Yau on which the topological string theory lives.
While this relation was conjectured to hold for compact Calabi-Yau manifolds, it is natural to expect it to continue to hold for non-compact manifolds as well. In fact, a concrete realization of this relation was given in the case of a non-compact Calabi-Yau in \cite{Vafa:2004qa}.

Therefore, one purpose of this paper is to try to better understand the relation between matrix models at special radii and the dual, non-compact, Calabi-Yau description. Because the Calabi-Yau is non-compact we expect divergences to appear in the calculations, thereby introducing cutoffs into the formalism. Since a natural goal is really to understand compact Calabi-Yau manifolds, we are careful in keeping the cutoff-dependent terms in order to understand where they come from on the matrix model side. We identify the infinities on both sides, and also discuss how finite matrix models lead to finite results for certain quantities on the Calabi-Yau side.

Given the identifications we make, we can, by using quite general arguments, show that the entropy of the 4D black hole is given by the canonical free energy of the matrix model, according to\footnote{Up to an overall normalization, which we will comment on in the next section.}
\begin{equation} \label{eq:bhfreeenergy}
S_{BH}=-\frac{F_{MM}}{T_{s}},
\end{equation}
where $T_{s}$ is the ``self-dual'' temperature in question.

For the particular case of the type 0 theories it is important to clarify which is the relevant radius to use. This we do using the relations between the 0A and 0B theories. Of these two cases, the deformed matrix model seems to be particularly interesting, partly because the chemical potential $\mu$ and the charge $q$ enter the theory very much in the same way as it does for the the topological string, or, given (\ref{eq:BHtop}), as it does for the BPS black hole.

We therefore focus our attention mainly on the deformed case. A subtlety one then needs to be careful with, is the fact that with the deformed potential the even wave functions of the matrix model become unphysical, and must be removed. We explicitly explain what this must correspond to on the Calabi-Yau side. Given these results we can precisely reproduce known matrix model quantities, using purely geometric techniques on the Calabi-Yau. Of particular importance, then, is to make sure that the Calabi-Yau calculation truly reproduces the known expressions of the grand canonical and canonical free energies of the deformed matrix model, which indeed it does.

This paper is organized as follows: In section 2 we present the general results on the relation between $4D$ black hole quantities and matrix model quantities. The observation that the black hole entropy is given in terms of the canonical free energy is explained. Particular emphasis is given to the role played by the Legendre transformation between different ensembles. Section 3 is devoted to a discussion on type 0 matrix models, in particular the relation between 0A and 0B matrix models and the role played by self-dual and T-dual radii in these theories. Section 4 gives a precise recipe for extracting Calabi-Yau from type 0 matrix models, with particular emphasis given to the deformed case. It is explained how the removal of the even wave functions of the matrix model is translated into the geometry of the Calabi-Yau. The resulting geometry is consistent with previous calculations of the geometry using the ground ring of the related 2-dimensional string theory \cite{Ita:2004yn}. In section 5 we test the ideas of section 2 using the results obtained in section 4 for the particular case of the deformed matrix model. The geometry resulting from the removal of the even states is discussed. We then go on and calculate the grand canonical and canonical free energy of the matrix model using purely geometric techniques on the Calabi-Yau in question, finding precise agreements with the known matrix model quantities. We conclude in section 6 with a list of open questions raised by our results. Appendix A discusses general aspects of non-compact Calabi-Yau manifolds. Appendix B explains the relation between the naive geometry corresponding to the deformed matrix model, where care has not been taken to include only the odd states, and the correct one, where the even states have been removed.

\bigskip

\section{Black hole physics from matrix models} \label{sec:bhmm}

\bigskip

\subsection{Black hole entropy from matrix models} \label{sec:bhthermo}

We start in this section by reviewing some of the main results of \cite{Ooguri:2004zv}. There an interesting relation was found between the ``thermodynamic'' entropy $S_{BH}(q_{I},p^{I})$ of the $\mathcal{N}=2$, $4D$ black hole (made up of electric $q_{I}$ and magnetic $p^{I}$ charges) and the free energy function $\mathcal{F}_{BH}(\phi^{I},p^{I})$, related to the partition function by
\begin{equation} \label{eq:blackholefe}
Z_{BH}(\phi^{I},p^{I})=\exp[\mathcal{F}_{BH}(\phi^{I},p^{I})]\equiv \sum_{q_{I}}\Omega_{BH}(q_{I},p^{I})e^{-\phi^{I}q_{I}}.
\end{equation} 
The potentials $\phi^{I}$ are the fixed potentials of the electric charges $q_{I}$, which are summed over in this ensemble \cite{Ooguri:2004zv}. The black hole entropy is then given by a Legendre transformation of the free energy according to \cite{Ooguri:2004zv}
\begin{equation} \label{eq:legendreentropy}
S_{BH}(q_{I},p^{I})=\mathcal{F}_{BH}(\phi^{I},p^{I})-\phi^{I} \frac{\partial}{\partial \phi^{I}}\mathcal{F}_{BH}(\phi^{I},p^{I}).
\end{equation}

For later purposes it will be useful to rephrase this in terms of geometry of the Calabi-Yau. The genus zero contribution to the entropy can be written as \cite{Ooguri:2004zv},
\be \label{eq:clentropy}
S_{BH,0}=\frac{i\pi}{4}\int \Omega \wedge \Omegabar,
\ee
where $\Omega$ is the holomorphic (3,0)-form\footnote{We refer the reader to appendix \ref{app:generalities} for details on the geometry of non-compact Calabi-Yau manifolds.} (not to be confused with the microcanonical partition function $\Omega_{BH}$). The genus zero entropy (\ref{eq:clentropy}) will now recieve corrections in essentially two ways \cite{LopesCardoso:1998wt,LopesCardoso:1999cv,LopesCardoso:1999ur,LopesCardoso:1999xn}. First of all, the tree-level prepotential, $F_{0}$, defined by (see appendix \ref{app:generalities} for notation)
\be
F_{0I}=\int_{B^{I}}\Omega,
\ee
is corrected from its classical value. More importantly for us, however, is another correction to (\ref{eq:clentropy}) of the form
\be \label{eq:entropycorrection}
\delta S_{BH}=\frac{\pi}{2} \text{Im} \left( z^{I}\frac{\partial}{\partial z^{I}}F-2F \right),
\ee
where the $z^{I}$ correspond to the A-cycle periods of $\Omega$. Notice that this correction is not sensitive to terms in $F(z^{I})$ being homogeneous of degree two in $z^{I}$. In general then, this correction term only sees higher order contributions to $F(z^{I})$. There is, however, an important exception to this rule which has to do with cutoff-dependent terms. This will be discussed in section \ref{sec:bhentropy}. But let us now return to (\ref{eq:legendreentropy}) and see what it means in terms of the matrix model.

From the point of view of the matrix model (\ref{eq:legendreentropy}) looks rather intriguing since it is precisely this type of transformation which takes us from the grand canonical (fixed chemical potential $\mu$) to the canonical (fixed fermion number $N$) free energy. So the goal of this section will be to understand this relation from the point of view of the matrix model at ``self-dual'' radius $R_{s}$ (with corresponding temperature $T_{s}=1/(2\pi R_{s})$).

The thing we will take advantage of in this section is the identification of the partition functions. As we explained, the grand canonical free energy of the matrix model should be the same as the one for the black hole, thus allowing us to make the identification\footnote{We should point our here that there may be a different power on the right hand side of (\ref{eq:partidentity}), since the normalization of the topological string free energy depends on the normalization of $\Omega$. One reason we believe this normalization is correct, however, is because in comparing the superstring matrix models, to the bosonic one, there is in a sense a doubling of the free energy, particularly explicit in the type $0B$ case. Therefore, the type 0 matrix models are, in terms of the partition functions, the square of the bosonic one. From \cite{Ghoshal:1995wm} we have, $Z_{bosonic}=Z_{top}$, suggesting $Z_{0A/0B}=|Z_{top}|^2$. Admittedly, this reasonling is somewhat speculative. In principle, one could calculate the normalization by carefully comparing our normalization of $\Omega$ to the one in \cite{Ooguri:2004zv}.}
\begin{equation} \label{eq:partidentity}
Z_{BH}(\phi^{I},p^{I})=Z_{MM}(\mu^{I},p^{I}),
\end{equation} 
where on the matrix model side we choose to call the potentials $\mu^{I}$, in order to conform with the standard notation. From now on we will furthermore drop the superscript on $\mu^{I}$ and call the magnetic charge\footnote{Strictly speaking, for the deformed matrix model, there is no clear distinction between electric and magnetic charges. In this context, however, it is natural to call $q$ magnetic charges.} $q$ in order to make contact with the explicit case of the deformed matrix model. The more general case will, however, always be understood.

On the matrix model side, at temperature $T$ and chemical potential $\mu$, we formally have
\begin{equation} \label{eq:grandmm}
Z_{MM}(\mu,q)=\sum_{N=0}^{\infty} Z_{MM}(q,N) e^{-\frac{\mu N}{T}}.
\end{equation}
So the sum goes over the number of fermions $N$ while $q$ is kept fixed. In the language of \cite{Ooguri:2004zv} we would call a partition function of this type a ``mixed'' partition function. It is mixed in the sense that $N$ is in the grand canonical ensemble, while $q$ is in the canonical one. Notice that in thinking about the $4D$ black hole we should rather think of the number of fermions $N$ as corresponding to the electric charges of the black hole and $q$ as being the magnetic ones.  

The free energy of this system, defined by $-F_{MM}/T=\ln Z_{MM}$, can be written in terms of thermodynamic variables as
\begin{equation}
-\frac{F_{MM}(\mu,q)}{T}=-\frac{E_{MM}}{T}+S_{MM} - \frac{\mu \langle N\rangle}{T},
\end{equation}
where $E_{MM}$ is the energy, $S_{MM}$ gives the entropy of the matrix model system, while $\langle N\rangle$ simply is the mean particle number. It will prove to be convenient now to transform to the canonical (or Helmholtz) free energy, according to
\begin{equation} \label{eq:legendre}
- F_{MM}(N,q)=- F_{MM}(\mu,q) +  \mu \frac{\partial}{\partial \mu} F_{MM}(\mu,q).
\end{equation}
Let us go to the self-dual radius $R_{s}$ where we expect the correspondence to work. The crucial observation now is that we are trying to calculate the entropy of a system at zero temperature (the extremal $D=4$, $\mathcal{N}=2$ black hole), using a model at non-zero temperature (the matrix model at $T=1/(2\pi R_{s})$). This means in particular that on the $4D$ black hole side, the energy $E_{BH}$ is zero (that is, the energy with respect to the ground state is zero, since the partition function only traces over ground state degeneracies). Thus on the $4D$ black hole side, we have
\begin{equation} \label{eq:bhfe1}
-\frac{1}{T_{BH}}{F_{BH}(N,q)}=S_{BH}(N,q),
\end{equation}
with the understanding that $T_{BH}=0$, while on the matrix model side we have
\begin{equation}
-\frac{F_{MM}(N,q)}{T_{s}}=-\frac{E_{MM}(N,q)}{T_{s}}+S_{MM}(N,q).
\end{equation}
From the identity $\ln Z_{MM}(N,q)=\ln Z_{BH}(N,q)$,
we therefore conclude
\begin{equation}
-\frac{F_{MM}(N,q)}{T_{s}}=-\frac{F_{BH}(N,q)}{T_{BH}}.
\end{equation}
Now, using (\ref{eq:bhfe1}), we finally obtain
\begin{equation} \label{eq:entropyformula}
S_{BH}(N,q)=-\frac{F_{MM}(N,q)}{T_{s}}.
\end{equation}
Eq. (\ref{eq:entropyformula}) tells us that the black hole entropy is given by the canonical free energy of the matrix model at self-dual radius, up to a proportionality constant given by minus the self-dual temperature.

In fact, notice that in comparing the respective grand canonical partition functions (\ref{eq:blackholefe}) and (\ref{eq:grandmm}) (at self-dual temperature), one is naturally led to the more general ``microscopic'' identification
\begin{equation} \label{eq:canonicalmm}
Z_{MM}(N,q)=\Omega_{BH}(N,q),
\end{equation}
which again is consistent with the fact that on the matrix model side we have a non-zero temperature while on the $4D$ black hole side the temperature is zero. That is, the partition function on the black hole side only calculates the number of states at the ground state, and so is nothing but the microcanonical partition function. 
\bigskip

\subsection{Matrix models and non-compact Calabi-Yau manifolds}
\label{sec:mmncCY}

\subsubsection{Finiteness of the canonical partition function}

Since we are working with non-compact Calabi-Yau manifolds, some of our
expressions will contain infinities which need to be regulated by introducing
a cutoff. On the other hand, since matrix models with sufficiently steep
potentials give finite results, we expect these infinities not to appear in
the geometric expressions that correspond to these results. In this section,
we will investigate how this comes about. A crucial point will turn out to be
that, contrary to the compact case, the prepotential of a noncompact
Calabi-Yau manifold is \emph{not} an ordinary function on its moduli space, but its form depends on a choice of coordinates.

Let us, for definiteness, consider a matrix model with a ``Mexican hat''
potential,
\begin{equation}
V(x) = - \frac{1}{2} x^{2} + \frac{1}{2} a^{2} x^{4}.
\end{equation}
As is well-known (see for example \cite{Aganagic:2003qj} for a very general
account of the relation between matrix models and topological strings), the
natural geometry related to such a matrix model with a fermi level $\mu$ is
the subspace of $\mathbb{C}^{4}$ defined by
\begin{equation}
uv + \mu= H(p,x) = \frac{1}{2} p^{2} - \frac{1}{2} x^{2} + \frac{1}{2} a^{2}
x^{4}.
\end{equation}
In fact we will see in sections 4 and \ref{sec:0ACY} that this
naive geometry is not always the correct one, but for our current purposes
these subtleties will not be important.

As we review in appendix \ref{app:generalities}, this Calabi-Yau has two
compact $A$-cycles and two noncompact $B$-cycles. As is also reviewed there,
the periods of the holomorphic three-form around these cycles are given by
\begin{align}
z^{I}  &  =  \int_{x_{-}^{I}}^{x_{+}^{I}} \sqrt{2 \mu+ x^{2} -
a^{2} x^{4}} \, dx\nonumber\\
w_{I}  &  =  \int_{x_{-}^{I}}^{\Lambda} \sqrt{2 \mu+ x^{2} - a^{2}
x^{4}} \, dx,
\end{align}
where $x_{\pm}^{I}$ are the boundaries of the $I$-th component of the
fermi sea, and $\Lambda$ is a large-distance cutoff for the noncompact
$B$-cycles. The compact periods $z^{I}$ only depend on $\mu$ and $a$. The
noncompact periods $w_{I}$ also depend on the cutoff $\Lambda$, but in a very
simple way. To see this, let us do a Taylor expansion:
\begin{align}
w_{I}  &  =  \int_{x_{-}^{I}}^{\Lambda} \sqrt{2 \mu+ x^{2} - a^{2}
x^{4}} \, dx\nonumber\\
&  =  \int_{x_{-}^{I}}^{\Lambda} \left(  i a x^{2} - \frac{i}{2a} +
O(x^{-2}) \right)  \, dx\nonumber\\
&  =  \frac{i a}{3} \Lambda^{3} - \frac{i}{2a} \Lambda+ \tilde{w}_{I}(\mu,a) +
O(\Lambda^{-1}),
\end{align}
with $\tilde{w}_{I}(\mu,a)$ independent of $\Lambda$. We see that, ignoring
terms which go to zero for large $\Lambda$, the $\Lambda$-dependent terms are
\emph{independent} of $\mu$; this will be crucial in what follows. Note that
any potential steeper than $x^{2}$ will lead to the same result, but that for
an $x^{2}$-potential we would get a $\mu\log\Lambda$ term.

As is well-known from special geometry (see appendix \ref{app:generalities}),
one can now define a prepotential $F_{0}(z)$ such that
\begin{equation}
w_{I} = \frac{\d F_{0}(z)}{\d z^{I}}.
\end{equation}
By expressing the $z^{I}$ in terms of $\mu$ and $a$, we can view this as a
function of $\mu$. Then the $\mu$-derivative of the prepotential is given by
\begin{equation}
\frac{\d F_{0}(\mu)}{\d\mu} = \frac{\d z^{I}}{\d\mu} w_{I},
\end{equation}
Integrating this equation and doing a partial integration, we find
\begin{align}
F_{0} (\mu)  &  =  \int^{\mu} d \mu^{\prime}\left(  \frac{\d z^{I}}{\d
\mu^{\prime}} w_{I} \right)  + c(a, \Lambda)\nonumber\\
&  =  z^{I}(\mu) w_{I}(\mu) - \int^{\mu} d \mu^{\prime}\left(  z^{I} \frac{\d
w_{I}}{\d\mu^{\prime}} \right)  + c^{\prime}(a, \Lambda)\nonumber\\
&  =  z^{I}(\mu) \left(  \frac{i a}{3} \Lambda^{3} - \frac{i}{2a}
\Lambda\right)  + f(\mu, a)\label{eq:wrongfe}%
\end{align}
Here, $f$ is an unknown but finite function of $\mu$ and $a$, and $c^{\prime}$
is a $\mu$-independent function. In the last line, we inserted $c^{\prime}=0$,
which can be seen to be true by the following argument. If we choose $\mu$
such that the fermi seas are exactly empty, both periods $z^{I}$ will vanish.
Moreover, we expect the prepotential to vanish in this limit. From the second
line in the above calculation, we then see that we can choose $c^{\prime}(a, \Lambda) = 0$.
We expect $c^{\prime}=0$ also for more general potentials, but we do not have
a complete proof of this fact. However, we can give the following argument. A
general polynomial of (even) degree $p$ has $p+1$ coefficients The coefficient
of the $x^{p}$ term should be fixed, because this determines the long distance
behaviour of the manifold, and hence the scale of $\Lambda$. Then, the $z^{I}$
can be written as functions of $p/2$ combinations of these coefficients. The
remaining $p/2$ degrees of freedom can be used to deform the potential without
changing the prepotential. One of these degrees of freedom corresponds to a
shift in the $x$-direction; it seems that the other $p/2-1$ give precisey
enough freedom to make sure that the $p/2$ minima of the potential are all at
the same height. Then for this resulting potential, the argument above can be
used to show that $c^{\prime}=0$.

The $\mu$- and cutoff-dependent term in (\ref{eq:wrongfe}) makes it hard to
identify this expression with the matrix model free energy, which is finite
without introducing any cutoffs. One gets a clue about how to solve this
problem by noting that this infinity comes from the partial integration we had
to do. However, suppose now we have a prepotential $\tilde{F}_{0}(w)$ which is
written in terms of the variables $w_{I}$. By definition,
\begin{equation}
z^{I} = \frac{\d\tilde{F}_{0}(w)}{\d w_{I}}.
\end{equation}
The $\mu$-derivative of this prepotential is
\begin{equation}
\frac{\d\tilde{F}_{0}(\mu)}{\d\mu} = \frac{\d w_{I}}{\d\mu} z^{I}.
\end{equation}
Note that this expression is completely finite! This means that the
prepotential
\begin{equation}
\tilde{F}_{0}(\mu) = \int^{\mu} d \mu^{\prime}\left(  \frac{\d w_{I}}{\d
\mu^{\prime}} z^{I} \right)  + \tilde{c}(a, \Lambda)
\end{equation}
at worst has an infinity which is independent of $\mu$. Even though in this
case we cannot use the argument which we used before, we would expect that
also this infinity is not present and the resulting expression is completely
finite. We propose that it is this prepotential (with the possible additive
infinity subtracted if necessary) that should be identified with the matrix
model partition function.

That this is a natural proposal can be seen by considering the simple case
where $V(x)=-\frac{1}{2}x^{2}$, which corresponds to the matrix model for the
$c=1$ string. In this case, one finds
\begin{align}
w  & \sim\Lambda^{2}+\mu\log(\mu/\Lambda^{2})\nonumber\\
z  & \sim\mu.
\end{align}
Here, we stick to the convention that the non-compact periods are called $w$. However, note that now these are the integrals over the fermi sea. The first expression above corresponds to what in the matrix model literature is
usually called $\Delta$. For matrix models, the partition function in terms of
$\Delta$ is a Legendre transform of the partition function in terms of $\mu$.
In other words, we have
\begin{equation}
\Delta=\frac{\partial F\left(  \mu\right)  }{\partial\mu},
\end{equation}
and%
\begin{equation}
\tilde{F}\left(  \Delta\right)  =\mu\frac{\partial F\left(  \mu\right)
}{\partial\mu}-F\left(  \mu\right)  .
\end{equation}
Using our results from the Calabi-Yau calculations, we find a grand canonical
partition function, with a divergent piece linear in $\mu$, given by%
\begin{equation}
F\left(  \mu\right)  \sim\Lambda^{2}\mu+\frac{\mu^{2}}{2}\log(\mu/\Lambda
^{2}).
\end{equation}
The canonical partition function, on the other hand, lacks such a contribution
(it is cancelled in the Legendre transform) and is given by%
\begin{equation}
\tilde{F}\left(  \Delta\right)  \sim\frac{\mu^{2}}{2}\log(\mu/\Lambda^{2}).
\end{equation}
The $\log\Lambda$ divergence common to the two expressions is easily regulated,
as in the example of the quartic potential, while the divergent term linear in
$\mu$ remains in $F\left(  \mu\right)  $.  In the canonical partition
function, on the other hand, the divergence is simply absorbed into a shift of
$\Delta$. Up to the $\log\Lambda$ divergence, $\tilde{F}\left(  \Delta\right)
$ is seen to be finite in terms of $\mu$.

Summarizing, we have two types of infinities. One is related to the infinite fermi sea, and can be regulated by modifying the potential. The other one is a consequence of the non-compactness of the Calabi-Yau, and it can be removed by using the right variables. As we have seen, thi sis similar to making a Legendre transform in the matrix model. Let us now see how the Legendre transform works in more detail from the
perspective of the Calabi-Yau.

\subsubsection{The Legendre transform}
One thing the reader might be surprised about after reading the previous section
is the following: how can the two prepotentials lead to numerically different
results? In special geometry, the prepotential is invariant under the interchange of the A- and the B-cycle periods. However, this is only true in the case of a compact
Calabi-Yau, where the prepotential is homogeneous of degree two. In general, the
prepotential $\tilde{F}_0(w)$ is a Legendre transform of the prepotential
$F_0(z)$. To show this, let us simply define $\tilde{F}_0(w)$ to be the Legendre
transform of $F_0(z)$:
\bea
 \tilde{F}_0(w) & = & z^I(w) \frac{\d F_0(z(w))}{\d z^I} - F_0(z(w) \ret
 & = & z^I(w) w_I - F_0(z(w)).
\eea
Now the $w_I$-derivative of this expression is
\bea
 \frac{\d \tilde{F_0(w)}}{\d w_I} & = & \frac{\d z(w)}{\d w_I} w_I + z^I(w) -
 \frac{\d z(w)}{\d w_I} \frac{\d F_0(z(w))}{\d w_I} \ret
 & = & \frac{\d z(w)}{\d w_I} w_I + z^I(w) - \frac{\d z(w)}{\d w_I} w_I \ret
 & = & z^I(w).
\eea
That is, $\tilde{F}_0(w)$ is indeed the prepotential in the coordinates $w_I$.
Note that when $F$ is homogeneous of degree two,
\be
 z^I \frac{\d F_0(z)}{\d z^I} = 2 F_0(z),
\ee
so we have
\be
 \tilde{F}_0(w) = F_0(z(w))
\ee
as required. The above results seem to suggest that we should really think of
$F_0$ as of a function which can be defined either in terms of ``velocities'' $z$ or of ``momenta'' $w$. This is 
reminiscent of the holomorphic anomaly equation of \cite{Bershadsky:1993cx} 
(see \cite{Witten:1993ed,Dijkgraaf:2002ac} for 
more on this intriguing subject). This equation tells us that the 
topological string partition function on a compact Calabi-Yau is not really 
a function of $z^I$, but it obtains a $\zbar^I$-dependence which turns the 
partition function into a function on a phase space defined by these 
variables. In particular, it has been suggested before\footnote{We thank 
A.~Neitzke for a discussion on this point.} that one might consider the 
Legendre transform to be the classical limit of the Fourier transform which 
lies at the base of the holomorphic anomaly. On the other hand, for the 
holomorphic anomaly the natural interpretation of the change of variables 
is as going from ``$p$'' to ``$q$'', whereas the Legendre transform is 
really a transformation between ``$p$'' and ``$\dot{q}$''. It would be 
interesting to work out the relation between the two transformations in 
more detail.

\subsubsection{The black hole entropy} \label{sec:bhentropy}

The next quantity we need to consider is the the black hole entropy. At genus
zero it was conjectured to be given by \cite{Ooguri:2004zv}
\begin{equation}
S_{BH,0}=\frac{i\pi}{4}\int\Omega\wedge\overline{\Omega}.
\end{equation}
Since we are working with a noncompact \textquotedblleft
compactification\textquotedblright\ manifold, we certainly do not expect this
result to be finite. In fact, by using the Riemann bilinear relation (see
appendix \ref{app:generalities}) we can write it as
\begin{equation}
S_{BH,0}=\frac{i\pi}{4} \left( z^{I}\overline{w}_{I}-\overline{z}^{I}w_{I} \right).
\label{eq:SBH}
\end{equation}
In the case of the $c=1$ model, this would be $\sim\Lambda^{2}\mu
+\mu^{2}\log(\mu/\Lambda^{2})$.
It is clear that the result does not depend on the choice of coordinates for our prepotential, and
moreover that it is indeed infinite if we remove the cutoff $\Lambda$.
However, as discussed in section \ref{sec:bhthermo}, there is a correction
term given by (\ref{eq:entropycorrection}) which should be added to yield the full entropy.
Contributions to $F\left(  \mu\right)  $ that are homogenous of degree two in
$\mu$ do not give any new contributions and one would, therefore, expect only
higher orders in the string coupling ($\sim1/\mu^{2}$) to be important. An
exception to this rule is, however, the cutoff dependent terms. In the
presence of these the correction term does give a non-vanishing contribution
that completes the entropy formula, and makes sure it is of the form of the
canonical partition function of the matrix model. That is,%
\[
S_{BH} \sim \mu^{2}\log(\mu/\Lambda^{2}).
\]
The only remaining cutoff dependence is due to the size of the Fermi sea and
is easily made finite as in our example with the quartic potential.  

In general of course, the Legendre transform with respect to $\mu$ will not
equal the Legendre transform with respect to all $w_{I}$, but at least the
fact that the result is infinite should not surprise us. In section
\ref{sec:0ACY} we will calculate all of the quantities mentioned above in the
case of the deformed (0A) matrix model, thus confirming the picture we have
sketched here.

An interesting side remark is that one can also calculate the Legendre
transform of (\ref{eq:SBH}) itself (with respect to both $z^{I}$ and
$\overline{z}^{I}$), leading to the expression
\begin{equation}
K(z, \overline{z}) = 2 z^{I} (\Im F_{IJ}) \overline{z}^{J}.
\end{equation}
As we can see from (\ref{eq:wrongfe}), this expression is manifestly finite.
It is the natural K\"ahler potential on the complex structure moduli space of
the Calabi-Yau. In the compact case, it would actually equal $i \int
\Omega\wedge\overline{\Omega}$. Here, however, this last expression is
infinite, but doing a Legendre transform we can extract a finite K\"ahler
potential from it.

\bigskip

\subsection{Density of states from the Calabi-Yau}
\label{sec:densityofstates}
A nice check on the relation $F_{MM}(\mu) = F_0(w_I(\mu))$ (we leave out the
tilde on the prepotential now) can be done by calculating the second
$\mu$-derivative of this equation. For the matrix model, it is well-known -- see
e.\ g.\ \cite{Klebanov:1991qa} -- that the result is the density of states:
\be
 \rho = \frac{\d^2 F_{MM}}{\d \mu^2}.
\ee
A semi classical approximation of the result can be obtained by the WKB-method,
and is given by
\be
 \rho = \int_{x_-}^{x_+} \frac{1}{\sqrt{2(\mu - V(x))}} d x,
 \label{eq:densityofstates}
\ee
where $x_{\pm}$ are the ``turning points'' where $V(x) = \mu$. Here, for
simplicity, we are restricting to the case that there is only a single
component of the fermi sea; the general expression will be a sum of terms of
this form.

We want to recover this same result from the topological string side at
genus zero -- that is, we want to calculate $\d^2 F_0 / \d \mu^2$. As we have
seen, the first derivative of $F_0$ with respect to $\mu$ is given by
\bea
 \frac{\d F_0}{\d \mu} & = & z^I \frac{\d w_I}{\d \mu}
 \label{eq:dftopdmu}
\eea
Again, let us focus on the contribution of a single component of the fermi sea
-- that is, of a single pair of $A$- and $B$-cycles. It will be useful to make
a change of symplectic basis of the three-cycles such that the dual cycle to
$A^I$ is $B_I - B_J$, where $B_J$ is an ``adjacent'' non-compact
cycle\footnote{Note that, even though this makes this B-period finite, for a
steep enough potential this is {\em not} necessary to make (\ref{eq:dftopdmu})
finite!}. The resulting B-cycle is compact, and the integration over it
corresponds to an ``under the hill'' integration in the matrix model.

Let us now make the extra assumption that for the new B-cycle integral, the
potential can be approximated by a quadratic potential. For example, this is
exactly true in the case where we have taken a double scaling limit in the
matrix model; it is also approximately true for general models when we take the
fermi sea to be ``nearly full''. In this case, the B-cycle integral will be
proportional to $\mu$, as can be easily verified, and hence (\ref{eq:dftopdmu})
is simply proportional to the $A$-cycle period $z^I$. Taking another
$\mu$-derivative of (\ref{eq:dftopdmu}) then gives
\bea
 \frac{\d^2 F_0}{\d \mu^2} & \sim & \frac{\d}{\d \mu} \int_{x_-}^{x_+} 
 \sqrt{2 (\mu - V(x))} dx \ret
 & = &
 \int_{x_-}^{x_+} \frac{1}{\sqrt{2 (\mu - V(x))}}.
\eea
Here, the differentiation under the integral sign is allowed because, even
though the boundaries depend on $\mu$, when we change $\mu$ slightly their
contribution is of order $(\gd \mu)^2$, since the integrand vanishes at the
boundary. Up to the overall prefactor which we could not determine from this
general argument (and which can be absorbed in a definition of $\Omega$ anyway)
this exactly coincides with (\ref{eq:densityofstates}). Of course, when the
potential is not steep enough, there will also be an infinite contribution from
the remaining non-compact B-cycle, as is also the case on the matrix model side.
\bigskip

\section{Matrix models for type 0 strings} \label{sec:mmtype0}

\bigskip

We will now turn to a more detailed description of the type $0A$ and type $0B$ strings. They are given by essentially the same matrix models
that were introduced more than 10 years ago in the context of the bosonic two
dimensional string, i.e., the $c=1$ model \cite{Gross:1990ay,Brezin:1989ss,Ginsparg:1990as}. For completeness, let us briefly
review the underlying principles of the matrix model.\footnote{An excellent
early review of the subject can be found in \cite{Klebanov:1991qa}.}

The main idea is to triangulate the string worldsheet using (dual) Feynman
diagrams of a quantum mechanical system based on $N\times N$ dimensional
hermitian matrices. The solution of the matrix model involves an integration
over the $N^{2}-N$ angular degrees of freedom leaving $N$ eigenvalues living
in a potential, which, in a limit we will describe below, is given by a
inverted harmonic oscillator potential,%
\begin{equation}
V\left(  x \right)  =-\frac{1}{2\alpha^{\prime}}x^{2}.
\end{equation}
The angular integration introduces a Vandermonde determinant that effectively
makes the eigenvalues fermionic. In this way we find a Fermi sea with a Fermi
level denoted by $\mu$, measuring the height from the top of the potential. To
obtain a theory of smooth string worldsheets one must take a limit -- the
double scaling limit -- where $N\rightarrow\infty$ and $\beta\rightarrow
\infty$, (where $\beta=1/\hslash$) in the appropriate way. More precisely, one
introduces $\Delta=\kappa_{c}^{2}-\frac{N}{\beta}$, where $\kappa_{c}^{2}$ is
a critical value determined by the precise form of the potential, and take
$\Delta\rightarrow0$. The free energy as a function of $\mu$ is related to the
free energy as a function of $\Delta$ through a Legendre function. In terms of
the Fermi level the double scaling limit requires that we take $\mu
\rightarrow0$ such that $\mu\beta$ remains finite. $1/\mu\beta$ plays the role
of the string coupling. Since $\mu\rightarrow0$ it is effectively only the
piece of the potential close to the top that is important for physics of
continuous surfaces.

While the matrix model essentially provides a nonperturbative definition of
the $c=1$ string, there was some confusion in the early literature about the
uniqueness of the nonperturbative extension. Filling just one side of the
potential gives a perturbatively stable situation, but due to tunneling there
are nonperturbative instabilities. There are a couple of ways to come to terms
with these instabilities. One can either fill the other side of the potential
to the same energy, or erect an unpenetrable wall at $x=0$. Neither of
these options affect the perturbative results, but the nonperturbative
corrections \textit{are} changed, which we will come back to later. \ An
unsatisfactory part of the story is, furthermore, what role, if any, the
existence of the other Fermi sea really plays. Is it there just for the
stability or does it represent another copy of the $c=1$ theory?

Another matrix model that was introduced in the early nineties in the context
of the bosonic string was the deformed matrix model, \cite{Jevicki:1993zg},
with potential given by%
\begin{equation}
V\left(  x \right)  =-\frac{1}{2\alpha^{\prime}}x^{2}+\frac
{M}{2 x^{2}}.
\end{equation}
The model was extensively studied on its own merits and also claimed to be
related to a two dimensional black hole, \cite{Danielsson:1993wq}%
-\cite{Danielsson:1994sk}. Interestingly, for values $M>\frac{3}{4}$, there
are no longer any non-perturbative instabilities in the model. The reason is
that we, in order to have normalizable wave functions, have no choice but to
pick states which are odd under reflection around $x=0$, i.e. we
effectively assume a wall sitting at $x=0$ where the wave functions vanish.

In the following section we will see how the the type 0A and type 0B string
theories fit into this framework.

\subsection{The 0B matrix model}

In \cite{Takayanagi:2003sm,Douglas:2003up} it was realized that the 0B string
theory can be described by a non-perturbatively stable matrix model based on
the inverted harmonic oscillator, i.e.,%
\begin{equation}
V\left(  x \right)  =-\frac{1}{4\alpha^{\prime}} x^{2},
\end{equation}
with both sides of the potential filled. The type 0B matrix model potential
only differs from the bosonic case through a simple factor of two. Contrary to
the case of the $c=1$ string, the two sides of the potential play a crucial
role in the type 0B theory. In the bosonic theory there is only one kind of
scalar perturbation, the (massless) tachyon, corresponding to ripples on a
single Fermi sea. In the case of type 0B strings, however, we have in addition
to the tachyon also a RR-scalar. In the matrix model the tachyon corresponds
to ripples which are even around $x=0$, while the RR-scalar corresponds
to odd ones.

The free energy of the type 0B matrix model is easily obtained from the old
matrix model through the appropriate identifications. We denote the free
energy at temperature $T$ by $F\left(  \mu_{B},R_{B}\right)  $, where
$R_{B}=\frac{1}{2\pi T}$ is the radius of compact, Euclidean time. By
convention we absorb $\beta$ into $\mu$. We then get
\begin{equation}
f\left(  \mu_{B},R_{B}\right)  =\frac{F\left(  \mu_{B},R_{B}\right)  }{T}=2\pi
RF\left(  \mu_{B},R_{B}\right)  ,
\end{equation}
where%
\begin{align}
f\left(  \mu_{B},R_{B}\right)   &  =2\operatorname{Re}\sum_{n,m=0}\ln\left(
\frac{1}{2\sqrt{2\alpha^{\prime}}}\left(  2n+1\right)  +\frac{2m+1}{2R_{B}%
}+i\mu_{B}\right) \label{eq:f0B}\\
&  =2\operatorname{Re}\sum_{n,m=0}\ln\left(  \frac{\sqrt{R_{B}}}%
{2\sqrt{2\alpha^{\prime}}}\left(  2n+1\right)  +\frac{2m+1}{2\sqrt{R_{B}}%
}+i\mu_{B}\sqrt{R_{B}}\right)  +\mathrm{const.}%
\end{align}
We will be focusing on universal and non-analytic contributions to the free
energy and can therefore ignore additive constants. To actually calculate the
free energy it is convenient, in practice, to compute the second derivative of
the free energy with respect to the Fermi level $\mu_{B}$, given by%
\begin{equation}
\frac{\partial^{2}f\left(  \mu_{B},R_{B}\right)  }{\partial\mu_{B}^{2}%
}=2\operatorname{Re}\sum_{n,m=0}\frac{1}{\left(  \frac{1}{2\sqrt
{2\alpha^{\prime}}}\left(  2l+1\right)  +\frac{2m+1}{2R_{B}}+i\mu_{B}\right)
^{2}},
\end{equation}
which also is the correlation function of two zero momentum tachyons. The free
energy, up to a couple of constants of integration, is then obtained by
integrating twice. In this way, the perturbative expansion in the string
coupling $\sim\frac{1}{\mu_{B}^{2}}$ of the free energy gives%
\begin{equation}
f\left(  \mu_{B},R_{B}\right)  =-\sqrt{2\alpha^{\prime}}R\mu_{B}^{2}\ln\mu
_{B}+\frac{1}{12}\left(  \frac{R_{B}}{\sqrt{2\alpha^{\prime}}}+\frac
{\sqrt{2\alpha^{\prime}}}{R_{B}}\right)  \ln\mu_{B}+...
\end{equation}

For the rest of our analysis it is useful to split the free energy into an odd
and an even part according to%
\begin{align}
f\left(  \mu_{B},R_{B}\right)   &  =2\operatorname{Re}\sum_{l,m=0}\ln\left(
\frac{1}{2\sqrt{2\alpha^{\prime}}}\left(  4l+3\right)  +\frac{2m+1}{2R_{B}%
}+i\mu_{B}\right) \nonumber\\
&  +2\operatorname{Re}\sum_{l,m=0}\ln\left(  \frac{1}{2\sqrt{2\alpha^{\prime}%
}}\left(  4l+1\right)  +\frac{2m+1}{2R_{B}}+i\mu_{B}\right)  .
\end{align}
That is, we write the free energy as%
\begin{equation}
f\left(  \mu_{B},R_{B}\right)  =f\left(  \mu_{B},R_{B}\right)  _{odd}+f\left(
\mu_{B},R_{B}\right)  _{even}, \label{eq:R/2}%
\end{equation}
where%
\begin{align}
f\left(  \mu_{B},R_{B}\right)  _{odd}  &  =f\left(  \frac{\mu_{B}+\frac
{i}{2\sqrt{2\alpha^{\prime}}}}{2},2R_{B}\right) \\
f\left(  \mu_{B},R_{B}\right)  _{even}  &  =f\left(  \frac{\mu_{B}-\frac
{i}{2\sqrt{2\alpha^{\prime}}}}{2},2R_{B}\right)  .
\end{align}
It is easy to see that the two terms have the same perturbative expansion but
differ non-perturbatively \cite{Danielsson:1992bx}. After all, the type 0B free energy of two Fermi
seas, $f\left(  \mu_{B},R_{B}\right)  $, is, perturbatively, simply twice the
free energy of a single sea, $f\left(  \mu_{B},R_{B}\right)  _{odd}$, where we
have put a wall at $x=0$. To see this in more detail, it is useful to
express the free energy in terms of the $\psi$-function, defined through
$\psi\left(  z\right)  =\frac{d\ln\Gamma\left(  z\right)  }{dz}$, which
formally can be written%
\begin{equation}
\psi\left(  z\right)  =-\sum_{n=0}\frac{1}{n+z}.
\end{equation}
If we focus on the case $R_{B}\rightarrow\infty$, we find%

\begin{align}
\lim_{R\rightarrow\infty}\frac{1}{R}\frac{\partial^{2}f\left(  \mu,R\right)
}{\partial\mu^{2}}  &  =2\operatorname{Re}\sum_{n=0}\left[  \frac{1}{\frac
{1}{2\sqrt{2\alpha^{\prime}}}\left(  4n+1\right)  +i\mu}+\frac{1}{\frac
{1}{2\sqrt{2\alpha^{\prime}}}\left(  4n+3\right)  +i\mu}\right] \\
&  =-2\sqrt{2\alpha^{\prime}}\operatorname{Re}\psi\left(  \frac{1+2\sqrt
{2\alpha^{\prime}}i\mu}{4}\right)  -2\sqrt{2\alpha^{\prime}}\operatorname{Re}%
\psi\left(  \frac{3+2\sqrt{2\alpha^{\prime}}i\mu}{4}\right)  .
\end{align}
Using that the $\psi$-functions obey
\begin{equation}
\psi\left(  1-z\right)  =\psi\left(  z\right)  +\pi\cot\pi z,
\end{equation}
and $\operatorname{Re}\psi\left(  z\right)  =\operatorname{Re}\psi\left(
\bar{z}\right)  $, we find
\begin{equation}
\operatorname{Re}\psi\left(  \frac{3+2\sqrt{2\alpha^{\prime}}i\mu}{4}\right)
=\operatorname{Re}\psi\left(  \frac{1+2\sqrt{2\alpha^{\prime}}i\mu}{4}\right)
+\frac{\pi}{\cosh\pi\sqrt{2\alpha^{\prime}}\mu}.
\end{equation}
As a consequence we find only nonperturbative differences between the odd and
even sums. Following \cite{Klebanov:1991ai}, the corresponding expression at
finite $R$ is easily obtained by applying the differential operator
\begin{equation}
\mathcal{R}=\frac{\frac{1}{R}\frac{\partial}{\partial\mu}}{e^{\frac{i}{2R_{{}%
}}\frac{\partial}{\partial\mu_{{}}}}-e^{-\frac{i}{2R}\frac{\partial}%
{\partial\mu_{{}}}}},
\end{equation}
using
\begin{equation}
e^{z\frac{\partial}{\partial x}}f\left(  x\right)  =f\left(  x+z\right)  .
\end{equation}

The expression for the free energy, separated into the odd and the even parts,
also suggests a way of obtaining the free energy at radius $R/2$, given the
result at radius $R$. In fact, the construction can be developed into a
procedure taking us from $R$ to $R/n$ using%
\begin{equation}
f\left(  \mu,\frac{R}{n}\right)  =\sum_{k=-\frac{n-1}{2}}^{\frac{n-1}{2}%
}f\left(  \frac{\mu+\frac{ik}{\sqrt{2\alpha^{\prime}}}}{n},R\right)  ,
\end{equation}
where (\ref{eq:R/2}) is just the special case of $n=2$ \cite{Gopakumar:1998vy}. The result immediately
follows from%
\begin{equation}
g\left(  n+\frac{1}{2}\right)  =\sum_{n}\sum_{k=-\frac{n-1}{2}}^{\frac{n-1}%
{2}}g\left(  p\left(  n+\frac{1}{2}\right)  +k\right),
\end{equation}
where $g$ is some arbitrary function. It is important to note that the function $f(b+ia, R)$ is invariant under sign changes of $b$ but {\it not} $a$.

We have now completed our preliminary analysis of the matrix model of the type
0B string, and we are ready to turn to the case of type 0A.

\subsection{The 0A matrix model}

The type 0A string is described, according to \cite{Douglas:2003up}, by a
deformed matrix model with potential
\begin{equation}
V\left(  \lambda\right)  =-\frac{1}{4\alpha^{\prime}} x^{2}+\frac
{M}{2 x^{2}},
\end{equation}
where%
\[
M=q^{2}-\frac{1}{4},
\]
and $q$ is a background RR-flux. Following \cite{Danielsson:1993wq}, we know
that we in the calculation of the free energy we are supposed to keep only the
odd states. The free energy then becomes
\begin{align}
f\left(  \mu_{A},R_{A},q\right)  _{def}  &  =2\operatorname{Re}\sum_{l,m=0}%
\ln\left(  \frac{1}{2\sqrt{2\alpha^{\prime}}}\left(  4l+3\right)  +\frac
{2m+1}{2R_{A}}+i\mu_{A}+\frac{-1+2\sqrt{M+\frac{1}{4}}}{2\sqrt{2\alpha
^{\prime}}}\right) \\
&  =2\operatorname{Re}\sum_{l,m=0}\ln\left(  \frac{1}{\sqrt{2\alpha^{\prime}}%
}\left(  2l+1\right)  +\frac{2m+1}{2R_{A}}+i\mu_{A}+\frac{\left\vert
q\right\vert }{\sqrt{2\alpha^{\prime}}}\right)  .
\end{align}
At $M=0$ (that is, $\left\vert q\right\vert =1/2$) we recover the odd part of
0B (at $R_{B}=R_{A}$), and in general we find%
\begin{equation}
f\left(  \mu_{A},R_{A},q\right)  _{def}=f\left(  \frac{\mu_{A}-\frac
{i\left\vert q\right\vert }{\sqrt{2\alpha^{\prime}}}}{2},2R_{A}\right)  .
\label{eq:fdef}%
\end{equation}
It is important here that there is only one term in the expression. Changing
$\left\vert q\right\vert \rightarrow-\left\vert q\right\vert $ does
\textit{not} give the same result as was explained in the
previous section.

\subsection{T-duality}

Let us now discuss how T-duality relates the 0A and the 0B strings\footnote{We focus, when going between $0A$ and $0B$, on the case $q=0$. A discussion of the interesting, and still not fully understood, case of $q\neq 0$ can be found in \cite{Gross:2003zz}.}. We begin
by noting that the 0B-expression (\ref{eq:f0B}) is (non-perturbatively)
self-dual under
\begin{equation}
R_{B}\rightarrow\frac{2\alpha^{\prime}}{R_{B}}, \label{eq:brdual}%
\end{equation}
provided that one rescales the Fermi level according to
\begin{equation}
\mu_{B}\rightarrow\frac{R_{B}}{\sqrt{2\alpha^{\prime}}}\mu_{B}.
\label{eq:bmudual}%
\end{equation}
Even more interesting, is the presence of another duality,%

\begin{align}
R_{A}  &  =\frac{\alpha^{\prime}}{R_{B}}\\
\mu_{A}  &  =\frac{\sqrt{2\alpha^{\prime}}}{R_{A}}\mu_{B},
\end{align}
which exchanges the 0A and 0B theories. When we apply these transformations to
the type 0B expression for the free energy, we indeed find
\begin{align}
f\left(  \mu,R_{B}\right)   &  =2\operatorname{Re}\sum_{n,m=0}\ln\left(
\frac{1}{2\sqrt{2\alpha^{\prime}}}\left(  2n+1\right)  +\frac{2m+1}{2R_{B}%
}+i\mu_{B}\right) \\
&  =2\operatorname{Re}\sum_{n,m=0}\ln\left(  \frac{1}{\sqrt{2\alpha^{\prime}}%
}\left(  2m+1\right)  +\frac{2n+1}{2R_{A}}+i\mu_{A}\right) \\
&  =f\left(  \mu_{A},R_{A},q=0\right)  _{def}=f\left(  \frac{\mu_{A}}%
{2},2R_{A}\right)  =f\left(  \frac{R_{B}}{\sqrt{2\alpha^{\prime}}}\mu
_{B},\frac{2\alpha^{\prime}}{R_{B}}\right)  .
\end{align}
In passing we may note that the last expression is equal to the first through
the 0B self duality. It is crucial to observe that even though the type 0B
expression includes both odd and even states the duality transformation
cleverly makes sure that the 0A expression only includes the odd ones as it should.

Since the 0B expression is non-perturbatively self dual around $R=$
$\sqrt{2\alpha^{\prime}}$, there will be a corresponding nonperturbative self
duality in type 0A. It is given by
\begin{align}
R_{A}  &  \rightarrow\frac{\alpha^{\prime}}{2R_{A}}\\
\mu_{A}  &  \rightarrow\sqrt{\frac{2}{\alpha^{\prime}}}R_{A}\mu_{A},
\label{eq:nydual}%
\end{align}
that is, the radius of exact self duality on the type 0A side is $\sqrt
{\frac{\alpha^{\prime}}{2}}$, which is just the $R_{A}=\frac{\alpha^{\prime}%
}{R_{B}}$ picture of the $R\rightarrow\frac{2\alpha^{\prime}}{R}$ self duality
of 0B. The self duality carries over to non-zero $q$ provided one also rescales%

\begin{equation}
q\rightarrow\sqrt{\frac{2}{\alpha^{\prime}}}R_{A}q.
\end{equation}
This is the self duality observed in \cite{Danielsson:1993dh} and further
discussed in \cite{Kapustin:2003hi}.\footnote{To compare with (\ref{eq:nydual}%
) one must take $\alpha^{\prime}\rightarrow2\alpha^{\prime}$ in
\cite{Danielsson:1993dh} to get the corresponding results for the type 0A
string.}

It is important to note that there is another perturbative duality inherited
from viewing the type 0A as a deformed and truncated type 0B. That is,
\begin{equation}
f\left(  \mu,R,q=1/2\right)  _{def}=f\left(  \mu,R\right)  _{odd}%
\end{equation}
This expression is perturbatively self dual around $R_{A}=$ $\sqrt
{2\alpha^{\prime}}$ (without rescaling of $q=1/2$).

\bigskip

\section{Calabi-Yau from matrix models} \label{sec:CYfromMM}

\bigskip

A convenient trick, introduced in \cite{Gross:1990ay}, when solving the matrix
model, is a continuation to a right side up harmonic oscillator using
$\alpha^{\prime}\rightarrow-\alpha^{\prime}$. As discussed in
\cite{Danielsson:1991jf}, one can also calculate correlation functions of
discrete states, including tachyons at discrete momenta, in this framework.
Important tools are the step operators%
\begin{align}
a &  =\frac{ip}{\sqrt{2}}+\frac{x}{2\sqrt{\alpha^{\prime}}}\\
a^{\dag} &  =-\frac{ip}{\sqrt{2}}+\frac{x}{2\sqrt{\alpha^{\prime}}},
\end{align}
obeying the algebra%
\begin{equation}
\left[  a,a^{\dag}\right]  =\frac{1}{\sqrt{2\alpha^{\prime}}},
\end{equation}
where we have suppressed Planck's constant $1/\beta$. The Hamiltonian is given
by
\begin{equation}
H=\frac{1}{2}\left(  a^{\dag}a+aa^{\dag}\right)  ,
\end{equation}
and tachyonic perturbations with positive momentum are obtained by acting with
$\left(  a^{\dag}\right)  ^{k}$, where $k$ is the momentum in units of
$\frac{1}{\sqrt{2\alpha^{\prime}}}$. Negative momentum tachyons are obtained
from acting with $a$.\footnote{In case of the type 0B string we should, to be
precise, distinguish between odd perturbations (RR-scalars) and even
perturbations (tachyons).}

While this construction works for any value of the radius, there exist further
simplifications, as explored in, e.g., \cite{Mukhi:1993zb}%
\cite{Ghoshal:1993qt}\cite{Hanany:1994fi}, which turn out to reproduce the
matrix model results at the self dual radius. To see this, we start with the
equation for the Fermi surface, $H=\mu$, which we formally write as%
\begin{equation}
a=\frac{\mu}{a^{\dag}}.
\end{equation}
If we want to calculate the correlation function between a single negative
momentum tachyon (momentum $-l$) and a number of positive momentum tachyons,
we therefore need to consider%
\begin{equation}
T_{-l}\sim a^{l}=\left(  \frac{\mu}{a^{\dag}}+\sum_{k=1}^{\infty}t_{k}\left(
a^{\dag}\right)  ^{k-1}\right)  ^{l}.\label{eq:Tundef}%
\end{equation}
The $t_{k}$ measures ripples on the Fermi sea corresponding to positive
momentum tachyons. The next step is to represent the algebra using $a^{\dag
}=\frac{1}{\sqrt{2\alpha^{\prime}}}\frac{\partial}{\partial X}\equiv D$, and
$a=\frac{\mu}{a^{\dag}}-X$, and to continue back  $\alpha^{\prime}%
\rightarrow-\alpha^{\prime}$ such that $D\rightarrow\frac{-i}{\sqrt
{2\alpha^{\prime}}}\frac{\partial}{\partial X}$. It then turns out that the
correlation functions at the self dual radius can be obtained from%
\begin{equation}
\left\langle T_{1}T_{-p}\right\rangle =\frac{1}{p}\mathrm{res}\left(
W^{p}\right)  ,\label{eq:t1tp}%
\end{equation}
where
\begin{equation}
W=\frac{\mu}{D}-X+\sum_{k=1}^{\infty}t_{k}D^{k-1},\label{eq:wc=1}%
\end{equation}
and the residue is defined as the coefficient of $1/D$. $X$ can also take the
role of $t_{1}$ and measure perturbations corresponding to the first discrete
tachyon, $T_{1}$. The expression (\ref{eq:t1tp}) is easily integrated to yield%
\begin{equation}
\left\langle T_{-p}\right\rangle =\frac{1}{p\left(  p+1\right)  }%
\mathrm{res}\left(  W^{p+1}\right)  .
\end{equation}
Keeping careful track of the ordering, as done in \cite{Danielsson:1994ac},
one can use these expressions to calculate correlation functions to all genus.
One only needs to recall that Planck's constant, $1/\beta$, has been absorbed
into the definition of $\mu$, and that tree level corresponds to the limit of
large $\mu$ (or rather large $\beta\mu$). 

Let us now generalize this construction to the deformed matrix model. As
explained in the previous section, the energy levels in the deformed matrix
model are shifted in a simple way as compared to the ordinary matrix model
based on the inverted harmonic oscillator. To be more precise, the odd and the
even levels (when present) shift in opposite directions. As a consequence
there is not a single kind of step operator but, instead, two types of step
operators given by%
\begin{equation}
a_{\pm}=\frac{ip}{\sqrt{2}}+\frac{x}{2\sqrt{\alpha^{\prime}}}\pm\frac{\sqrt
{M}}{\sqrt{2}x},
\end{equation}
which go between odd and even levels respectively. We also have operators%
\begin{align}
b &  =\left(  \frac{ip}{\sqrt{2}}+\frac{x}{2\sqrt{\alpha^{\prime}}}\right)
^{2}-\frac{M}{2x^{2}}\\
b^{\dag} &  =\left(  -\frac{ip}{\sqrt{2}}+\frac{x}{2\sqrt{\alpha^{\prime}}%
}\right)  ^{2}-\frac{M}{2x^{2}},
\end{align}
which take two steps at a time and go from even to even or from odd to odd.
Since, when $M$ or $q$ is large enough, we must project out the even levels,
we need to work with $b$ and $b^{\dag}$, rather than with the $a_{\pm}$ and
$a_{\pm}^{\dag}$ which take us out of the Hilbert space. Using the appropriate
operators we can reconstruct the Hamiltonian as (up to operator ordering)%
\begin{equation}
b^{\dag}b=H^{2}-\frac{M}{2\alpha^{\prime}}\rightarrow H^{2}+\frac{M}%
{2\alpha^{\prime}},
\end{equation}
where we have continued back to the upside down case by taking $\alpha
^{\prime}\rightarrow-\alpha^{\prime}$. It is now easy to write down the analog
of (\ref{eq:Tundef}), which simply is given by%
\begin{equation}
T_{-2l}\sim b^{l}=\left(  \frac{1}{b^{\dag}}\left(  \left(  \mu+\sum
t_{2k}\left(  b^{\dag}\right)  ^{k}\right)  ^{2}+\frac{M}{2\alpha^{\prime}%
}\right)  \right)  ^{l}.
\end{equation}
Following the procedure in the case of the undeformed matrix model, we now
need to represent the algebra between the $b$ and the $b^{\dag}$, that is%
\begin{equation}
\left[  b^{\dag},b\right]  =\frac{4i}{\sqrt{2\alpha^{\prime}}}H,
\end{equation}
(for the upside down case) in a convenient way. Note that the algebra of the
$b$ and the $b^{\dag}$ coincides with the algebra between $a^{2}$ and $\left(
a^{\dag}\right)  ^{2}$ in the undeformed matrix model. It is easy to verify
that%
\begin{align}
b^{\dag} &  =D^{2}\\
b &  =\frac{1}{D^{2}}\left(  \left(  \mu+XD\right)  ^{2}+\frac{M}%
{2\alpha^{\prime}}\right)  ,
\end{align}
with $D=\frac{-i}{\sqrt{2\alpha^{\prime}}}\frac{\partial}{\partial X}$, does
the job. That is, one finds%
\begin{equation}
\left[  b^{\dag},b\right]  =\frac{4i}{\sqrt{2\alpha^{\prime}}}\left(
\mu+XD\right)  =\frac{4i}{\sqrt{2\alpha^{\prime}}}H.
\end{equation}
The above structure is exactly what was used in \cite{Danielsson:1994ac} to
calculate correlations functions, including higher genus, at $R=\sqrt
{2\alpha^{\prime}}$. There it was argued, up to the shift $\alpha^{\prime
}\rightarrow2\alpha^{\prime}$ which takes us between the bosonic and type 0
strings, that one should work with%

\begin{equation}
W=\frac{M}{2\alpha^{\prime}D^{2}}+\left(  \frac{\mu}{D}-X+\sum t_{2k}%
D^{2k-1}\right)  ^{2},
\end{equation}
in order to obtain the correlation functions. In our heuristic derivation we
have not paid attention to the precise ordering of the operators. In
\cite{Danielsson:1994ac}, however, it is shown how the above expression yields
the correct correlation functions also at higher genus.

Let us now discuss how the results we have obtained so far is used to define
Calabi-Yau conifolds as discussed earlier. In the undeformed case, the
conifold equation is simply obtained from
\begin{equation}
H-\mu=a^{\dag}a-\mu
\end{equation}
by writing%
\begin{equation}
a^{\dag}a-\mu=uv. \label{eq:c=1con}%
\end{equation}
In the deformed case we have%
\begin{equation}
b^{\dag}b=\mu^{2}+\frac{M}{2\alpha^{\prime}}%
\end{equation}
and it is natural to write%
\begin{equation}
b^{\dag}b=\left(  \mu+uv\right)  ^{2}+\frac{M}{2\alpha^{\prime}}.
\label{eq:bbcon}%
\end{equation}
The expressions that go into the calculation of the correlation functions of
the previous section, are recovered when we restrict the conifold equation to
the Fermi surface represented by $uv=0$. In the next section we will
investigate in more detail how this works from the point of view of the Calabi-Yau.

It is important to contrast this with what happens when we use (\ref{eq:R/2})
to go to half the self dual radius for the undeformed matrix model. The free
energy is, in this case, a sum over two terms, and the conifold equation
becomes%
\begin{align}
uv &  =\left(  \frac{\mu_{B}+\frac{i}{2\sqrt{2\alpha^{\prime}}}}{2}-a^{\dag
}a\right)  \left(  \frac{\mu_{B}-\frac{i}{2\sqrt{2\alpha^{\prime}}}}%
{2}-a^{\dag}a\right)  \\
&  =\left(  \frac{\mu_{B}}{2}-a^{\dag}a\right)  ^{2}+\frac{1}{8\alpha^{\prime
}}.\label{eq:R/2con}%
\end{align}
The two expressions (\ref{eq:bbcon}) and (\ref{eq:R/2con}) look superficially
similar but there are some important differences. The deformed matrix model
essentially skips half of the terms in the sum over $n$ in (\ref{eq:f0B}),
while going to half the self dual radius essentially means skipping half the
terms in the sum over $m$. The two operations are more or less T-dual to each
other with the role of winding and momentum interchanged. Or, in other words,
an interchange of $a^{\dag}a$ and $uv$ as is apparent from the two conifold equations.

The differences between the two cases become even more clear, if we use the
conifold equation for the case of half the selfdual radius to derive the
expression corresponding to (\ref{eq:wc=1}). To do this, we must solve for
$a^{\dag}a$. This yields two solutions given by%
\begin{equation}
a^{\dag}a=\frac{\mu_{B}}{2}\pm\sqrt{uv-\frac{1}{8\alpha^{\prime}}}.
\end{equation}
At the Fermi sea, $uv=0$, we find%
\begin{equation}
a^{\dag}a=\frac{\mu_{B}}{2}\pm\frac{i}{2\sqrt{2\alpha^{\prime}}},
\end{equation}
which translates into a sum over two contributions with $\frac{\mu_{B}}{2}$
shifted $\pm\frac{i}{2\sqrt{2\alpha^{\prime}}}$. It is easy to verify that
this indeed gives the right answer for tachyon correlation functions even at
higher genus.

Finally, we need to make contact with the expression for the type 0A free
energy that we derived in the previous section, (\ref{eq:fdef}). Can we obtain
the conifold equation directly as we did above for the case of half the self
dual radius? To do this we use the selfduality of $f\left(  \mu,R\right)  $ to
write (\ref{eq:fdef}) as%
\begin{equation}
f\left(  \mu_{A},R_{A},q\right)  _{def}=f\left(  \frac{\mu_{A}-\frac
{i\left\vert q\right\vert }{\sqrt{2\alpha^{\prime}}}}{2},2R_{A}\right)
=f\left(  \frac{R_{A}}{\sqrt{2\alpha^{\prime}}}\left(  \mu_{A}-\frac
{i\left\vert q\right\vert }{\sqrt{2\alpha^{\prime}}}\right)  ,\frac
{\alpha^{\prime}}{R_{A}}\right)  .
\end{equation}
Evaluating this at $R_{A}=$ $\sqrt{2\alpha^{\prime}}$ we find%
\begin{align}
f\left(  \mu_{A},R_{A},q\right)  _{def}=f\left(  \mu_{A}-\frac{i\left\vert
q\right\vert }{\sqrt{2\alpha^{\prime}}},\sqrt{\frac{\alpha}{2}^{\prime}%
}\right) & = f\left(  \frac{\mu_{A}-\frac{i\left\vert q\right\vert }%
{\sqrt{2\alpha^{\prime}}}+\frac{i}{2\sqrt{2\alpha^{\prime}}}}{2},\sqrt
{2\alpha^{\prime}}\right) \\ & +f\left(  \frac{\mu_{A}-\frac{i\left\vert
q\right\vert }{\sqrt{2\alpha^{\prime}}}-\frac{i}{2\sqrt{2\alpha^{\prime}}}}%
{2},\sqrt{2\alpha^{\prime}}\right)  .
\end{align}
That is, a structure of the same form as (\ref{eq:R/2}) and consistent with
the similarity between (\ref{eq:bbcon}) and (\ref{eq:R/2con}). To reconstruct
the confold equation from here, it is important to realize that there is an
ambiguity due to the invariance of the free energy with respect to the sign of
$\mu$. Keeping terms leading in Planck's constant $1/\beta$, it is easy to
convince one self that, indeed, (\ref{eq:bbcon}) is a conifold consistent with
the matrix model results. While the difference between writing $M$ or $q^{2}$
in, e.g., (\ref{eq:bbcon}), disappears when we reinsert $\beta$ and take
$\beta\rightarrow\infty$, it is the expression with $M$, as shown in
\cite{Danielsson:1994ac}, which carries over to calculations to higher genus.

It is interesting to note that the construction suggests a way of calculating
correlation functions involving tachyons with momentum as well as winding. The
key is to consider changes in the complex structure depending on \textit{all}
the complex coordinates on the conifold. Specifically, if we consider the case
of the bosonic or type 0B theory, we would consider adding perturbations
depending not only on $a$ and $a^{\dag}$ (corresponding to tachyons with
momentum) but also on $u$ and $v$ (corresponding to tachyons carrying winding).

\bigskip

\section{Calculation of the type 0A results from the CY} \label{sec:0ACY}

\bigskip

Using the knowledge we obtained in the previous section, we now want to test
the ideas of section \ref{sec:bhmm} in the particular case of the deformed
matrix model. Before turning to the calculations, however, there is a seeming
paradox we have to resolve. Naively, one would expect the topological string
theory corresponding to the deformed matrix model to have a target space of the
form
\be
 X: \qquad uv + \mu = H(p,x) = \half p^2 - \half x^2 + \frac{M}{2 x^2},
 \label{eq:Xdefeqn}
\ee
where $H$ is the Hamiltonian of the type 0A matrix model. However, whenever the
matrix model under consideration corresponds to a two-dimensional string
theory, as is the case here, there is another natural geometry to consider.
This geometry is given by the defining equations of the so-called ground ring
\cite{Witten:1991zd}. In the most studied case, the one of the $c=1$ string at
selfdual radius, both approaches lead to exactly the same deformed conifold
geometry, but in general the two geometries may be different.

For the deformed matrix model, when we consider the ground ring of the
corresponding string theory, as was done in detail in \cite{Ita:2004yn} for the
case of charge $q=0$, one indeed finds a manifold which differs from
(\ref{eq:Xdefeqn}). In \cite{Ita:2004yn}, a proposal for the defining equation
of the ground ring in the case $q \neq 0$ was also made. We recovered this
equation from a somewhat different perspective, and up to a slight change in
the coefficients, in section \ref{sec:mmtype0}. For reasons which will become
clear in a moment, let us denote this manifold by $X'/2$:
\be
 X'/2: \qquad (uv + \mu)^2 = b b^\dagger - M
 \label{eq:Xp2defeqn}
\ee
Note that even though we use the notation in terms of $b$ and $b^\dagger$ from
section \ref{sec:CYfromMM}, here we consider $b$ and $b^\dagger$ to be ordinary
complex variables, not operators\footnote{However, as was explained in
\cite{Aganagic:2003qj}, the correct interpretation of the $(b,
b^\dagger)$-plane is that of a phase space, which gets quantum corrections
that make $b$ and $b^\dagger$ noncommutative. It would be very interesting to
work this out in more detail for this example.}.

At first, one might hope that both of the above equations lead to the same
manifold, but a simple study of for example their singularities shows that this
is not the case. Why then is it not correct to work with $X$? The reason is
that in defining the deformed matrix model, we have restricted our space of
states to only the odd ones. This is a subtlety that the naive construction of
$X$ does not take into account. Therefore, one would expect hat this manifold
is correct ``up to a factor of two'', in a sense. In appendix
\ref{app:twomanifolds}, we make this statement precise by showing how one can
go from $X$ to $X'/2$ by a series of $\bZ_2$-orbifolds: first one divides out a
$\bZ_2$-symmetry of $X$ to arrive at a manifold $X/2$, and then one constructs
a different manifold $X'$ of which both $X/2$ and $X'/2$ are $\bZ_2$-orbifolds.

Let us now calculate the periods of $\gO$ on $X'/2$. This manifold is not of
the type considered in appendix \ref{app:generalities}, but the techniques used
to calculate the periods are quite similar. We view the geometry as a fibration
over the $(u,v)$-plane. Then, the fiber is a cylinder, except over the curves
$C_{\pm}: uv = -\mu \mp i \sqrt{M}$, where the one-cycle on the cylinder
pinches. (The change in sign in the notation is to make the result below look
simpler.) The geometry has two $A$-cycles, which we denote by $A_{\pm}$, and
which can be constructed by taking disks $D_{\pm}$ in the $(u,v)$-space which
have the non contractible circles on $C_{\pm}$ as their boundaries, and taking a
circle-fibration over these disks which degenerates into a point (the node of
$uv=0$) at the boundary of $D_{\pm}$. Then we immediately find\footnote{Up to
the shift by $1/4$ which we mentioned before, these periods agree with the
results in \cite{Ita:2004yn}.}:
\bea
 z^{\pm} = \int_{A_{\pm}} \gO & = &  \int_{A_{\pm}} \frac{du \wedge dv
 \wedge db}{b} \ret
 & = & 2 \pi i \int_{D_{\pm}} du \wedge dv \ret
 & = & 2 \pi i \int_{D_{\pm}} d (u dv) \ret
 & = & 2 \pi i \int_{\d D_{\pm}} u dv \ret
 & = & 2 \pi i \int_{\d D_{\pm}} \frac{-\mu \mp i \sqrt{M}}{v} dv \ret
 & = & 4 \pi^2 (\mu \pm i \sqrt{M})
 \label{eq:Aperiods}
\eea
Next, we calculate the $B$-periods of $\gO$ on $X'/2$. Because of the
non-compactness of these cycles, we really have to parameterize them, and
introduce a cutoff. Let us define a mapping from the upper half plane 
into $(u,v)$-space by
\bea
 u & = & c \left[ e^r + \frac{\theta}{\pi} \left( e^{-r} - e^r \right) \right]
 \ret
 v & = & c \left[ e^{-r} + \frac{\theta}{\pi} \left( e^r - e^{-r} \right)
 \right],
\eea
where $c^2 = -\mu \mp i \sqrt{M}$ and we parameterized the upper half plane by a
radial coordinate $r$ and an angular coordinate $\theta$ running from $0$ to
$\pi$. Note that the boundary of the image indeed lies on the curve $uv = c^2$.
We will integrate over the upper half plane up to some cutoff radius $R_c$. We
should take this cutoff $c$-dependent for the following reason. Note that the
last ``half-circle'' over which we integrate becomes the line
\bea
 u & = & c e^{R_c} \left( 1 - \frac{\theta}{\pi} \right) \ret
 v & = & c e^{R_c} \frac{\theta}{\pi}
\eea
in $(u,v)$-space, where we ignored terms of order $e^{-R_c}$. This
``target-space'' cutoff should be independent of $c$, and therefore we have to
choose
\be
 R_c = \log (\gL / c)
\ee
with $\gL$ large. Using this, the integral is straightforward to perform, and
we find
\be
 w_+ = \int_{B_+} \gO = i \pi \gL^2 - 2 \pi i \mu_c \log (\mu_c / \gL^2),
 \label{eq:Bperiods}
\ee
where we used the notation $\mu_c = \mu + i \sqrt{M}$. The result for
$w_-$ is given by the same expression with $\mu_c \to \mubar_c$. Note
that, as is also known from the case of the conifold for example, the
multivaluedness of the logarithm corresponds exactly to adding one or more
$A$-cycles to the $B$-cycle.

Now, let us calculate the prepotential of the geometry. Of course, $z_{\pm}$
are really complex conjugate variables, so it might seem that we do not have
sufficient information to do so, but we can solve this problem by formally
viewing $\mu$ and $M$ as complex variables, so $z_{\pm}$ become independent.
Integrating up the $B$-cycle periods to get the prepotential in terms of the
$z_{\pm}$ is then a simple exercise, and we find
\begin{equation}
F_{0}\left(  \mu,\sqrt{M}\right)  =\pi\Lambda^{2}z_{+}-\frac{z_{+}^{2}}{\pi
}\log\left(  \frac{z_{+}}{2\pi i \Lambda^{2}e^{2}}\right)  +(z_{+}%
\leftrightarrow z_{-}).\label{eq:fmum}%
\end{equation}
In this equations and the ones that follow, we are implicitly taking the imaginary part on the right hand side to find the real results from the matrix model literature. The above expression is the generalization of the grand canonical partition function for the
undeformed matrix model, $F\left(  \mu\right)  $, which we discussed earlier,
and has the caracteristic divergent term linear in $z$. As we explained in
section 2.2, it is natural to make a Legendre transform in all
the variables $z$ to obtain
\begin{equation}
\tilde{F}_{0}\left(  \Delta,p\right)  =\frac{\mu_{c}^{2}}{32\pi}\sqrt
{\frac{2}{\alpha^{\prime}}}\log\left(  \frac{\mu_{c}^{2}}{\Lambda^{4}}\right)
+(\mu_{c}\leftrightarrow\overline{\mu}_{c})\sim (\mu^{2}-M^{2}%
)\log\left(  \frac{\mu^{2}+M^{2}}{\Lambda^{4}}\right)  ,
\end{equation}
where we now changed the norm of $\Omega$ (and hence $z_{\pm}$), and we
absorbed the constant in the logarithm into $\Lambda$. $\tilde{F}_{0}$ is by
nature a function of $\Delta$ and the Legendre conjugate to $\sqrt{M}$, which
we have denoted by $p$. Up to the term linear in $\mu$, and after redefining $\gL$, this coincides,
numerically, with $F_{0}$. We may also choose to Legendre transform only in
$\mu$, and leave $\sqrt{M}$ as it is. In this case we find%
\begin{equation}
\breve{F}_{0}\left(  \Delta,\sqrt{M}\right)  =-\frac{1}{8\pi}(\mu^{2}%
+M^{2})\log\left(  \frac{\mu^{2}+M^{2}}{\Lambda^{4}}\right)
.\label{eq:fdeltam}%
\end{equation}
where we set $\alpha^{\prime}=1/2$. Note that indeed, our result agrees precisely with matrix model results, at
$\mu=0$, discussed recently in \cite{Danielsson:2003yi,Gukov:2003yp,Davis:2004xb,Danielsson:2004xf}.
Also note the crucial relative sign between $\mu$ and $M$, comparing (\ref{eq:fdeltam}) and (\ref{eq:fmum}).

Finally, let us calculate the genus zero contibution to the black hole entropy (\ref{eq:clentropy}). Using the Riemann
bilinear relation, one finds
\begin{align} \label{eq:entropycy}
S_{BH, 0}&=\frac{i\pi}{4}\int\Omega \wedge \Omegabar= \frac{i\pi}{4}\left(z_{+}\overline{w}_{+}+z_{-}\overline{w}_{-}\right)+\mbox{c.c.} \\ & =-\frac{1}{8}\mu\Lambda^{2}+\frac{1}{8}\mu_{c}\overline{\mu}_{c}\log(\mu
_{c}\overline{\mu}_{c}/\Lambda^{4})\\
& =-\frac{1}{8}\mu\Lambda^{2}+\frac{1}{8}(\mu^{2}+M^{2}%
)\log\left(  \frac{\mu^{2}+M^{2}}{\Lambda^{4}}\right)  ,\nonumber
\end{align}
Just as in the case of the undeformed
matrix model, this expression is modified by the correction term (\ref{eq:entropycorrection}) discussed in
section \ref{sec:bhthermo}, which only gets a contribution from the term linear in $\mu$. Correcting (\ref{eq:entropycy}) by (\ref{eq:entropycorrection}) and then comparing with (\ref{eq:fdeltam}), we find that $S_{BH, 0}=-\pi \breve{F}_{0}$. Since the self-dual temperature for the deformed matrix model is $T_{s}=1/\pi$, this is precisely eqn. (\ref{eq:entropyformula}) found in section 2.1.

\bigskip

\section{Conclusion}

\bigskip

In this paper we have studied the connection between $c=1$ matrix models at self-dual radii and non-compact Calabi-Yau manifolds. These results are particularly interesting in light of the recent proposal made in \cite{Ooguri:2004zv}, where the partition function of topological string theory was conjectured to be equivalent to the one for $\mathcal{N}=2$, $D=4$ black holes.

Using the findings of \cite{Ooguri:2004zv} and the connection between topological string theory and matrix models, formulated in this paper, we have calculated the entropy of the $4D$ black hole (\ref{eq:entropyformula}). As explained in the paper, the main observation is that the black hole entropy is related to the canonical free energy of the matrix model by 
\begin{equation}
S_{BH}=-\frac{F_{MM}}{T_{s}}.
\end{equation}
The reason this is the case is that on the space-time side the temperature of the black hole is zero, implying that the canonical partition function reduces to just the number of ground states of the theory.

In particular the relation between the deformed matrix model and Calabi-Yau has been analyzed in detail. In this construction it was important to clarify what one means with self-dual radius since, unlike the bosonic case, there seem to be many radii to choose among. We discussed this in detail in section 3, where the relation between 0B and 0A was used in order to clarify this issue. The second thing of great importance was to carefully take into account the fact that only the odd wave functions on the matrix model side should be included. This restriction naturally introduced the two-step operators $b$ and $b^{\dagger}$ into the geometry of the Calabi-Yau.

Given this we were able to precisely calculate matrix model quantities such as the grand canonical and canonical free energies purely from the Calabi-Yau side, reproducing the known matrix model results. Even though on the Calabi-Yau side we were working mainly at the classical level, we believe the generalization to a quantum description should be quite straightforward. In particular, the two-step operators $b$ and $b^{\dagger}$ explicitly present on the Calabi-Yau geometry should make this very interesting generalization possible.

To conclude this paper, we list a few open questions raised by our findings.
\begin{itemize}
	\item
	One obvious goal is to find a matrix model description of a compact Calabi-Yau. Since we don't know how to do this, in this paper we had to deal with non-compact Calabi-Yau manifolds and the resulting divergencies. Even though we found a good way to deal with this, it would be very interesting to see how the manifestly finite results from a compact Calabi-Yau can be translated into matrix model results or vice versa.
	\item
	The relation between matrix models and extremal $4D$ black holes is very interesting from the point of view of the $AdS_{2}$ near horizon geometry of the black holes, as was pointed out in \cite{Ooguri:2004zv,Vafa:2004qa}. On general grounds one would expect to find a $CFT_{1}$ dual of the system, i.e. a matrix quantum mechanics. One puzzling feature with the $AdS_{2}$ geometry is that it has two boundaries, suggesting that perhaps one should look for two holographic duals. At this point we do not have much to say about this apart from the potentially interesting observation that the deformed matrix model we consider in this paper necessarily has a reality condition put on it, effectively factorizing the theory into two chiral sectors. One might speculate on that they are naturally associated to each boundary respectively \cite{Vafa:2004qa}.
	\item
	In deriving the relation between the deformed matrix model and the Calabi-Yau, it was important to make sure to remove half of the eigenstates relative to the undeformed case. In the original relation between $2D$ string theory and $c=1$ matrix models, the removal of half of the states corresponds, for example, on the space-time side to projecting the RR-scalar away \cite{Takayanagi:2003sm,Douglas:2003up}, thus leaving the tachyon as the only propagating degree of freedom. It would be interesting to see if something similar happens on the $4D$ space-time side.
	\item
	As already mentioned, the generalization of our analysis to the quantum case would quite naturally be very interesting. For the specific case of the deformed matrix model it is quite clear that a very useful tool for this generalization are the two-step operators, explicitly sitting as parts of the geometry of the Calabi-Yau. Indeed, considering the commutation relations of the generators, it seems natural to expect some kind of non-commutative geometry to arise in this process. Probably the techniques of \cite{Aganagic:2003qj} will be very useful in this regard.
\end{itemize}

\bigskip

\section*{Acknowledgments}

UD is a Royal Swedish Academy of Sciences Research Fellow supported by a grant
from the Knut and Alice Wallenberg Foundation. The work was supported by
the Swedish Research Council (VR) and the Royal Swedish Academy of Science.  	   
 
\bigskip

\appendix
\section{Generalities on non-compact Calabi-Yau manifolds} \label{app:generalities}

\bigskip

In this paper, we consider non-compact complex three-dimensional Calabi-Yau
manifolds embedded in $\bC^4$ as the set of solutions of a defining equation
\be
 G(p,x,u,v) = 0.
 \label{eq:definingeqn}
\ee
The object we will be interested in most is the holomorphic (3,0)-form $\gO$
living on this hypersurface. Such a three-form is only well-defined up to
multiplication by a holomorphic function. On a compact Calabi-Yau, the only
globally defined holomorphic functions are the constant ones, so in that case
the holomorphic three-form is unique up to multiplication by a constant. On a
non-compact manifold, there is in general much more freedom in the choice of the
three-form\footnote{To avoid confusion, note that this does not mean that the
cohomology class (and hence periods) of the three-form are not unique!}. In the case where the three-fold is embedded in $\bC^4$, however, there
is a very natural choice for $\gO$. It follows from the natural choice for the
holomorphic four-form in the embedding space:
\be
 \gL^{(4)} = dp \wedge dx \wedge du \wedge dv
\ee
We would like to decompose this as $\gL^{(4)} = \gO \wedge A^{(1)}$, where
$A^{(1)}$ is a one-form perpendicular to the embedded hypersurface. (This means
that for a vector field Y along the hypersurface, $A^{(1)} \circ Y=0$.) The
natural choice for $A^{(1)}$ is of course $A^{(1)} = dG$.

So, for example, if in a certain patch we choose $(p,x,u)$ as local coordinates
on the three-dimensional surface, we would write
\be
 \gL^{(4)} = \left( \frac{dx \wedge dp \wedge du}{\d G / \d v} \right) \wedge
 \left( \frac{\d G}{\d v} dv \right),
\ee
and hence
\be
 \gO = \frac{dp \wedge dx \wedge du}{\d G / \d v},
\ee
where we have to express $\d G / \d v$ in terms of $p, x$ and $u$.

An important ingredient in our calculations will be the Riemann bilinear
relation, which says that for two closed three-forms $C_1^{(3)}$ and
$C_2^{(3)}$,
\be
 \int_{CY} C_1^{(3)} \wedge C_2^{(3)} = \sum_{I=1}^{b_3} \left( \int_{A^I}
 C_1^{(3)} \int_{B_I} C_2^{(3)} - \int_{A^I} C_2^{(3)} \int_{B_I} C_1^{(3)}
 \right),
 \label{eq:riemannbi}
\ee
where $b_3$ is the third Betti number of the Calabi-Yau and $A^I, B_J$ is a
canonical basis of three-cycles for $H_3(X)$, meaning that 
\bea
 A^I \cap B_J & = & \gd^I_J \ret
 A^I \cap A^J & = & 0 \ret
 B_I \cap B_J & = & 0.
 \label{eq:intersections}
\eea
Strictly speaking, the Riemann bilinear identity only holds for compact
Calabi-Yau spaces. For example, on a non-compact manifold a basis of the above
form may not even exist. However, even though our Calabi-Yau manifolds are
non-compact, we will always be thinking of them as patches of a larger, compact
Calabi-Yau. With this in mind, we will identify the ``physical'' $A$- and
$B$-cycles, and calculate the contributions to (\ref{eq:riemannbi}) coming from
these cycles.

The canonical $A$- and $B$-cycles can be used to find coordinates on the moduli
space of complex structures on the Calabi-Yau manifold. Given the cohomology
class of the holomorphic three-form $\gO$, such a complex structure is uniquely
defined. Since this cohomology class of a three-form is uniquely determined by
its periods around a basis of three-cycles, we can use these periods as a set of
coordinates on the moduli space of complex structures. It turns out that these
coordinates are highly redundant, and it is enough to specify only the periods
around the A-cycles:
\be
 z^I = \int_{A^I} \Omega.
\ee
In fact, there is still a slight redundancy, since multiplying $\Omega$ by a
complex constant does not change the complex structure. Therefore, the $z^I$ are
{\em projective} coordinates on the moduli space of complex structures of the
Calabi-Yau. In these coordinates, the B-cycle periods of $\Omega$ also have a
very simple expression:
\be
 \int_{B_I} \Omega = F_I \equiv \frac{\d F_0}{\d z^I},
\ee
where the prepotential $F_0$ can be calculated as the genus zero contribution to
the B-model topological string partition function $F_{top}$. As is conventional,
we will denote derivatives of $F_0$ with respect to the $z^I$ by $F_I, F_{IJ}$,
and so on.

Again, the above is only strictly true for a compact Calabi-Yau manifold. This
will be sufficient for our purposes, because of the assumed ``hidden
compactness'' we mentioned before. One thing which will be different from the
well-known compact properties, however, is that the prepotential is no
longer a homogeneous function of degree two in the $z^I$; there will be
logarithmic corrections to this coming from the fact that some of the
three-cycles are non-compact. As a result, the prepotential becomes dependent
on our choice of coordinates on the complex structure moduli space. This fact
plays an important role in section \ref{sec:mmncCY}.

In the case where we have a matrix model\footnote{In this paper, we use
``matrix model'' to mean ``matrix quantum mechanics'', although many statements
are also likely to be valid for ``ordinary'' matrix models where the matrices
do not depend on time.} with a Hamiltonian
\be
 H(p,x) = \half p^2 + V(x),
\ee
a very natural geometry to consider (see \cite{Aganagic:2003qj} for details) is
the one where the defining equation (\ref{eq:definingeqn}) takes the form
\be
 uv + \mu = H(p,x) = \half p^2 + V(x).
 \label{eq:generalgeometry}
\ee
The simplest example of this is the case when $V(x) = - \half x^2$, where the
corresponding geometry is the deformed conifold. As is well-known
\cite{Ghoshal:1995wm}, the topological string theory on the deformed conifold
indeed has a partition function which equals the grand canonical partition
function of a matrix model with potential $V$, namely the matrix model for the
bosonic $c=1$ string at selfdual radius.

\begin{figure}[ht]
 \begin{center}
  \includegraphics[height=5cm]{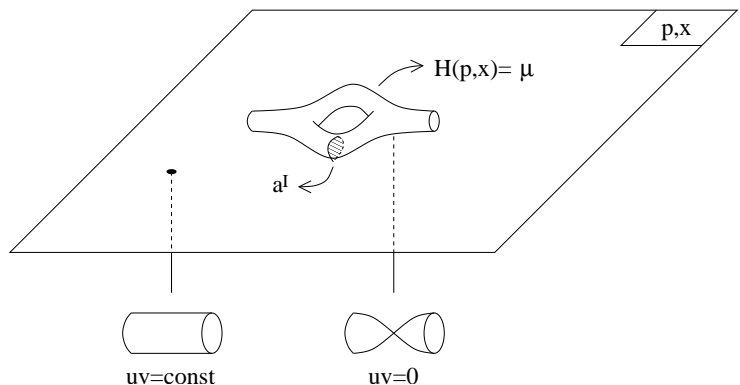}
 \end{center}
 \caption{Geometries of the form (\ref{eq:generalgeometry}) can be viewed as
 fibrations over the $(p,x)$-plane. When $H(p,x)=\mu$, the fiber degenerates.
 Cycles of the three-fold are constructed by taking cycles of this
 Riemann-surface, filling them up, and taking a fibration over them which is a
 circle at the interior and the pinched circle, i.\ e.\ a point, at the
 boundary.} 
 \label{fig:fibration}
\end{figure}

Let us now review the way in which one calculates the periods of $\Omega$ in a
geometry of the form (\ref{eq:generalgeometry}). One views the geometry as a
(degenerate) fibration over the complex $(p,x)$-plane, see figure
\ref{fig:fibration}. At a general point $(p,x)$, the fiber will be of the form 
$uv =$ {\em const}, which is topologically a cylinder. However, when $H(p,x) =
\mu$, the fiber degenerates into $uv = 0$, which is the union of two complex
planes identified in one point, or equivalently a cylinder with a pinched cycle.
The $A$- and $B$-cycles of the complex three-fold are now constructed by taking
$a$- and $b$-cycles of the Riemann surface $H(p,x) = \mu$, ``filling up'' these
cycles in the complex $(p,x)$-plane, and taking a circle fibration over them
which degenerates at the boundary on the Riemann surface. For compact one-cycles
on the Riemann surface, this leads to 3-cycles with the topology of a
three-sphere. For non-compact one-cycles, one needs to introduce a cutoff on the
Riemann surface; the corresponding three-dimensional topology is that of a
three-ball, where the $S^2$ at its boundary corresponds to the cutoff on the
Riemann surface. With some brain gymnastics, one can convince oneself that the
cycles constructed in this way are indeed non contractible and independent, have
the correct intersection properties, and that these are all the three-cycles.

\begin{figure}[ht]
 \begin{center}
  \includegraphics[height=3.5cm]{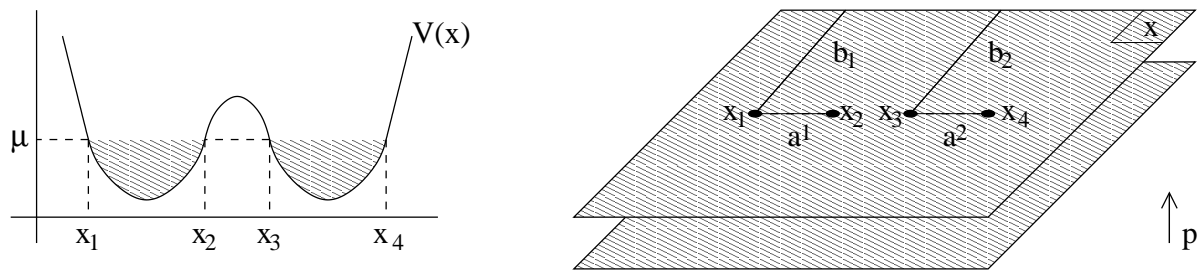}
 \end{center}
 \caption{The Riemann surface $\half p^2=V(x) - \mu$ consists of two sheets,
 glued together at a finite number of branch points $x_i$. The $a^I$ denote both
 the branch cuts and the canonical choice of $a$-cycles; the $b_I$ denote the
 canonical $b$-cycles (but no branch cuts).} 
 \label{fig:riemannsurface}
\end{figure}

To carry out the above procedure, we thus need to find the one-cycles on the
non-compact Riemann surface given by
\be
 \Sigma: \qquad \half p^2 + V(x) - \mu = 0.
\ee
Let us again view this surface as a fibration, this time over the complex
$x$-plane. At a general $x$, there will be two values of $p$ satisfying the
equation, so the fiber consists of two points. However, when $V(x) = \mu$ there
will be only a single point; let us call the points where this happens $x_i$.
Moreover, there will be branch cuts starting from the $x_i$: when walking around
$x_i$ in the $x$-plane once, one returns on the opposite $p$-sheet. This is
depicted in figure \ref{fig:riemannsurface}, where we chose a potential and a
$\mu$ such that all $x_i$ are real. Moreover, for convenience we chose an even
number of branch points. Note that the position of the branch cuts is of course
a choice; any choice of branch cuts which either end up at infinity or at
another branch point will do.

One can now construct non contractible one-cycles by going from $x_i$ to $x_j$
on one sheet, and go back on the other sheet. Moreover, one can go from any of
these points to infinity along both sheets, to get a non-compact cycle. The
conventional choice is to take all $a$-cycles to be compact and along the
branch-cuts, corresponding to integrals over the fermi sea components in the
matrix model, and to take the $b$-cycles going to infinity. One may be 
tempted to draw as many finite cycles as possible by connecting points which
are not connected by branch cuts, such as $x_2$ and $x_3$ in the figure, but
then the compact $b$-cycles will intersect two $a$-cycles and hence will not
satisfy the definition (\ref{eq:intersections}). However, such a compact cycle
can be viewed as the {\em difference} of two $b$-cycles\footnote{One can also
view $b_1 - b_2$ as a true $b$-cycle, provided one takes one of the $a$-cycles
to be $a_1 + a_2$ so it does not intersect this $b$-cycle.}, a point which will
be important to us in section \ref{sec:densityofstates}. Of course no matter
what, one is always going to end up with at least one truly non-compact cycle.

The periods of the three-form are now easily calculated by first integrating
over the fiber in the complex $(u,v)$-plane using Cauchy's theorem, and then
integrating over the two-dimensional surface in the base space, which we will
call $D$. This last integral is carried out by first integrating over the line
from a point $(p,x)$ to the point $(-p,x)$ using Stokes' theorem, and finally
integrating over the cut (for an $A$-cycle) or line to infinity (for a
$B$-cycle) in the $x$-plane. This gives:
\bea
 \int_{A/B} \gO & = & \int_{A/B} \frac{dx \wedge dp \wedge du}{u} \ret
 & = & 2 \pi i \int_D dp \wedge dx \ret
 & = & 2 \pi i \int_D d (p dx) \ret
 & = & 4 \pi i \int_{a/b} p(x) dx \ret
 & = & 4 \pi i \int_{a/b} \sqrt{2 (\mu - V(x))} dx.
\eea

\bigskip

\section{Relation between the two manifolds for 0A} \label{app:twomanifolds}

\bigskip

\subsection{Relating the manifolds}
In section \ref{sec:0ACY}, we explained that the naive manifold on which a
topological string dual to the deformed matrix model should live,
\be
 X: \qquad uv + \mu = \half p^2 - \half x^2 + \frac{M}{2 x^2},
 \label{eq:Xdefeqn2}
\ee
is not the correct one, but that it should be replaced by
\be
 X'/2: \qquad (uv + \mu)^2 = b b^\dagger - M
 \label{eq:Xp2defeqn2}
\ee
From the construction in section \ref{sec:mmtype0}, it is quite clear how the
defining equations of $X$ and $X'/2$ are related. Let us first square the
equation for $X$:
\be
 (uv + \mu)^2 = \frac{1}{4} (p^2 - x^2 + \frac{M}{x^2})^2.
 \label{eq:squaredeqn}
\ee
Then, as in section \ref{sec:mmtype0}, we make the following definitions of $b$
and $b^\dagger$:
\bea
 b & = & \half (p+x)^2 + \frac{M}{2 x^2} \ret
 b^\dagger & = & \half (p - x)^2 + \frac{M}{2 x^2}.
\eea
Inserting these definitions in (\ref{eq:Xp2defeqn2}), we recover exactly the
equation (\ref{eq:squaredeqn}).

Let us now see what this means in terms of the geometries. It is clear that the
map from $(p,x)$ to $(b, b^\dagger)$ is not one-to-one. One sees immediately
that $(p,x)$ and $(-p, -x)$ map to the same point, and with a short calculation
one finds that
\be
 \left( \frac{p x^2}{\sqrt{p^2 x^2 + M}}, \frac{\sqrt{p^2 x^2 + M}}{x} \right)
 \qquad \mbox{and} \qquad \left( - \frac{p x^2}{\sqrt{p^2 x^2 + M}}, -
 \frac{\sqrt{p^2 x^2 + M}}{x} \right)
\ee
also map to the same $(b, b^\dagger)$.

\begin{figure}[ht]
 \begin{center}
  \includegraphics[height=5cm]{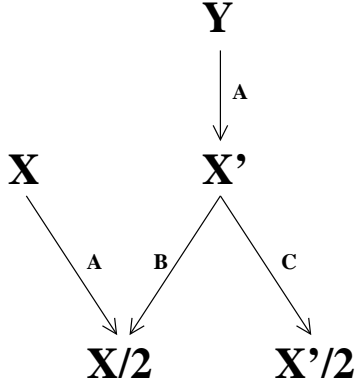}
 \end{center}
 \caption{The relations between the different manifolds introduced in this
 section.} 
 \label{fig:allmanifolds}
\end{figure}

With these results, it is easy to construct $X'/2$ out of $X$. Let us first
construct a manifold $X/2$ by dividing out the $\bZ_2$-action $(p,x) \mapsto
(-p, -x)$ on the manifold $X$:
\be
 X/2: \qquad \left( uv + \mu = \half p^2 - \half x^2 + \frac{M}{2 x^2} \right)
 / \bZ_2^A,
\ee
where we denoted this particular $\bZ_2$ by $\bZ_2^A$. Now, note that this
manifold is also a $\bZ_2$-orbifold of the following orbifold: 
\be
 X': \qquad \left( (uv + \mu)^2 = \left( \half p^2 - \half x^2 + \frac{M}{2
 x^2} \right)^2 \right) / \bZ_2^A.
\ee
That $X/2$ is a $\bZ_2$-orbifold of this can be easily seen by writing it as
\be
 X': \qquad \left( uv + \mu = \pm \left( \half p^2 - \half x^2 + \frac{M}{2
 x^2} \right) \right) / \bZ_2^A,
\ee
and noting that $(p,x,u,v) \mapsto (ip,ix,u,v)$ maps solutions with a plus sign
on the right hand side to solutions with a minus sign, and squares to the
identity. (It squares to minus the identity, which equals the identity because
this is exactly the $\bZ_2^A$ we divided out.) We denote this orbifolding
group by $\bZ_2^B$. Finally, to get from $X'$ to to $X'/2$, we need to divide
out another $\bZ_2$, which we call $\bZ_2^C$ and which maps\footnote{This map has appeared before in \cite{Takayanagi:2004ge}, where the 
author constructs a type IIA matrix model from a matrix model similar to 
our type 0A model. It would be interesting to see if one can rephrase the 
results of \cite{Takayanagi:2004ge} in terms of Calabi-Yau manifolds as well.}
\be
 (p,x) \mapsto \left(\frac{p x^2}{\sqrt{p^2 x^2 + M}}, \frac{\sqrt{p^2 x^2 +
 M}}{x} \right).
 \label{eq:Z2C}
\ee
Note that this is a well-defined $\bZ_2$-action only after dividing out
$\bZ_2^A: (p,x) \sim (-p, -x)$, since the square of this operation is $\pm 1$.
However, by a slight abuse of notation we will write
\be
 X'/2: \qquad \left( (uv + \mu)^2 = \left( \half p^2 - \half x^2 + \frac{M}{2
 x^2} \right)^2 \right) / (\bZ_2^A \times \bZ_2^C),
\ee
for the resulting manifold. We have summarized the different manifolds and their
relations in figure \ref{fig:allmanifolds}, where for clarity we also included
the manifold
\be
 Y: \qquad \left( (uv + \mu)^2 = \left( \half p^2 - \half x^2 + \frac{M}{2
 x^2} \right)^2 \right)
\ee
This manifold will not play any role in the rest of our story.

Summarizing, we see that the fact that in the matrix model we only have the odd
level states, translates into the chain of $\bZ_2$-orbifolds we encountered
above. In a sense, the resulting manifold $X'/2$ in indeed ``half as big'' as
the original manifold $X$. It would however be interesting to see a matrix model
interpretation of the several other manifolds in this chain as well. Roughly,
the picture seems to be as follows. When we go from $X$ to $Y$, we add a second
matrix model to the original one. Then $\bZ_2^A$ removes half of the states of
each matrix model, and $\bZ_2^C$ relates the two matrix models in such a way
that their moduli become complex conjugates. To get a better understanding of
the relation between matrix models and geometries, it would be interesting to
work out this structure in more detail and see if it can be applied to different
examples as well.

\subsection{Relating the cycles and periods}

It is instructive to see how the different three-cycles, and hence the
$\gO$-periods, in the different geometries are related. Since in the definition
of $X$, the explicit Hamiltonian of the matrix model appears, the physical
interpretation of the different cycles in terms of integrations over the fermi
sea or ``under the hill'' is much clearer there. Therefore, we would like to
see how these cycles are related to those on $X'/2$. Instead of parameterizing
the different cycles and calculating how they behave under the different
$\bZ_2$-maps, we choose to do this in a more graphical and as a result
hopefully more intuitive way.

\begin{figure}[ht]
 \begin{center}
  \includegraphics[height=5cm]{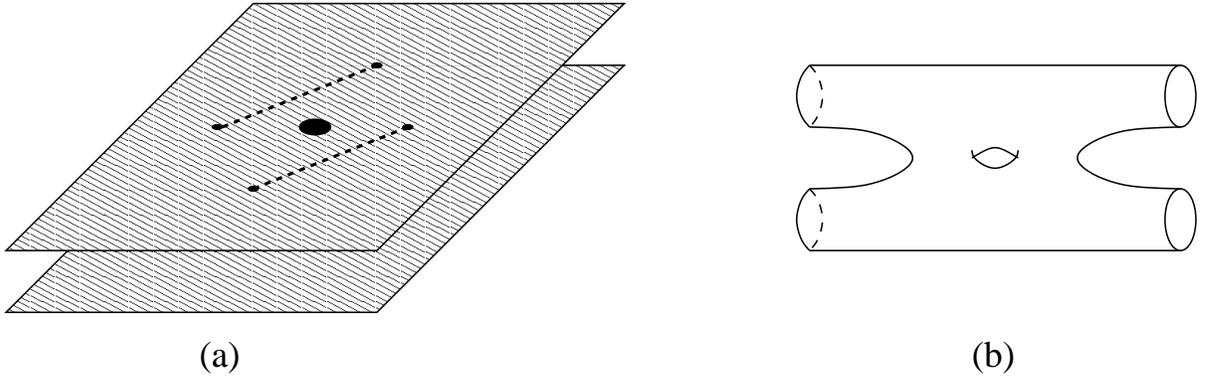}
 \end{center}
 \caption{(a) The leafs $\Ltilde_c$ consist of two sheets, connected at four
 branch points. The small dots indicate the branch points, and the dashed lines
 a choice of branch cuts. The big dot in the middle corresponds to $x=0$, and is
 not part of the sheets. (b) A smooth picture of the same topology.}
 \label{fig:bigleaf}
\end{figure}

We will use the same strategy as before, and think of all the geometries as
fibrations over the complex $(p,x)$-plane. Moreover, it will be useful to think
of this complex $(p,x)$-plane as foliated by leafs of the form
\be
 \Ltilde_c: \qquad \half p^2 - \half x^2 + \frac{M}{2 x^2} = c.
\ee
Just as before, we draw these leafs as degenerate fibrations over the complex
$x$-plane -- see figure \ref{fig:bigleaf}a. By viewing the two sheets in this
picture as cylinders, and the branch cuts as tubes joining these cylinders,  one
sees that the topology of these leafs is that of figure \ref{fig:bigleaf}b. When
$c = \pm i \sqrt{M}$, the two branch cuts degenerate into nodes. These
degenerate leafs will not play any role in what follows; they do not correspond
to special regions in the manifold, but their appearance is simply a result of
our way of foliating the space.

\begin{figure}[ht]
 \begin{center}
  \includegraphics[height=5cm]{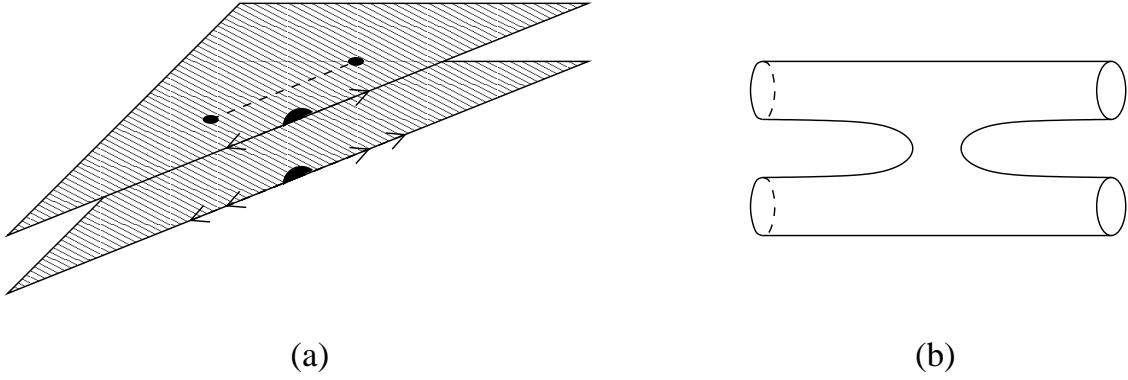}
 \end{center}
 \caption{(a) The leafs $L'_c$, which are $\bZ_2^A$-orbifolds of the leafs
 $\Ltilde_c$. The diagonal boundaries are identified as indicated by the arrows.
 (b) A smooth picture of the same topology.} 
 \label{fig:mediumleaf}
\end{figure}

In the manifolds where we have divided out the $\bZ_2^A$, mapping $(p,x)$ to
$(-p, -x)$, the leafs will be $\bZ_2$-orbifolds of $\Ltilde_c$, which we denote
by $L'_c$:
\be
 L'_c: \qquad \left( \half p^2 - \half x^2 + \frac{M}{2 x^2} = c \right) /
 \bZ_2^A.
\ee
We draw these leafs in figure \ref{fig:mediumleaf}a. Note that the boundaries of
the two triangles are identified within the {\em same} sheet, perhaps contrary
to what one would expect if one takes the picture too literally. That this is
the right identification can be most easily seen by noting that near the
singularity, the surface is given by $p = \pm i \sqrt{M} x^{-1}$. By gluing the
edges one sees that the topology is that of two cylinders connected by a single
tube, as in figure \ref{fig:mediumleaf}b.

\begin{figure}[ht]
 \begin{center}
  \includegraphics[height=5cm]{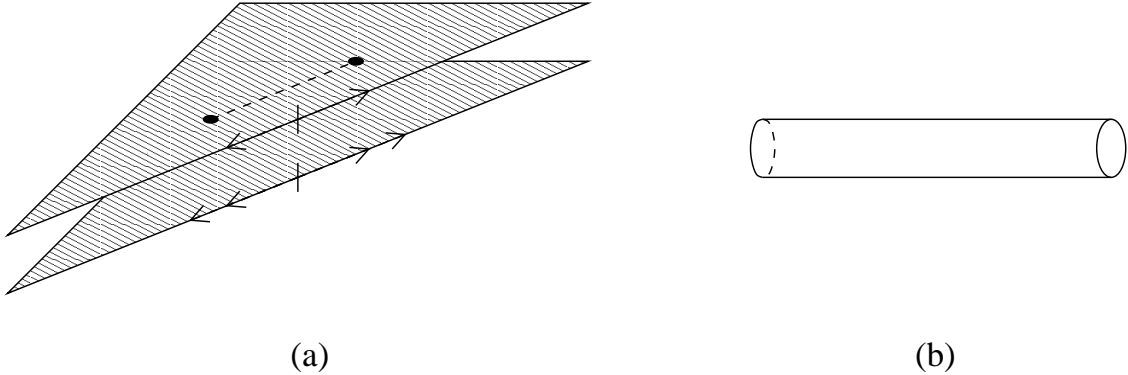}
 \end{center}
 \caption{(a) The leafs $L_c$. The diagonal boundaries are identified as
 indicated by the arrows. Note that the points $x=0$ have now become regular
 points. (b) The topology of $L_c$ is simply that of a cylinder.} 
 \label{fig:smallleaf}
\end{figure}

Finally, the identification $\bZ_2^C$ identifies points in the leaf $L'_c$ with
points in the leaf $L'_{-c}$. (Note that in the leaf $L'_0$, both elements of
$\bZ_2^C$ act as the identity.) We denote the resulting leafs by $L_c$:
\be
 L_c = (L'_c \cup L'_{-c}) / \bZ_2^C
\ee
At first sight, this does not seem to change the topology of the leafs, but note
that (\ref{eq:Z2C}) maps the two regular points with $x = i \sqrt{2c}$ in $L'_c$
(where $p^2 x^2 = -M$) to the two infinities at $x=0$ in $L'_{-c}$, and vice
versa. Therefore, after the identification the two points with $x=0$ become
regular points, as is drawn in figure \ref{fig:smallleaf}a. This leaf has the
topology of a cylinder, as was to be expected since it can alternatively be
written as $b b^{\dagger} = c + M$.

\begin{figure}[ht]
 \begin{center}
  \includegraphics[height=5cm]{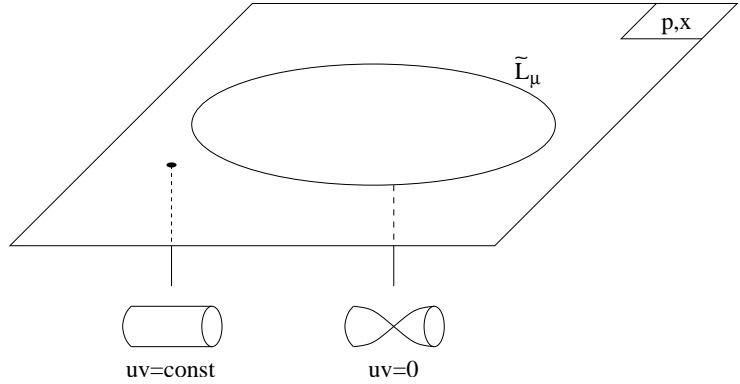}
 \end{center}
 \caption{The fibration giving the manifold $X$. At the leaf $\Ltilde_{\mu}$,
 the fiber degenerates. The same picture is valid for $X/2$, with the leaf
 $\Ltilde_{\mu}$ replaced by $L'_\mu$.} 
 \label{fig:fibrationX}
\end{figure}

Now let us consider the different fibrations. Let us start with
\be
 X: \qquad uv + \mu = \half p^2 - \half x^2 + \frac{M}{2 x^2}.
\ee
Here, the fiber over a general leaf $\Ltilde_c$ is of the form $uv = c - \mu$,
which is topologically a cylinder. Only when $c = \mu$, the fiber degenerates
into a pinched cylinder. This is drawn in figure \ref{fig:fibrationX}. Using
the same construction as before, we can construct several three-cycles in the
full geometry from one-cycles on the Riemann surface in figure
\ref{fig:bigleaf}. The one-cycles that will be of interest to us are the
compact $a$-cycles going between two branch points, and the non-compact
$b$-cycles going off to infinity from a single branch point. The physical
interpretation is the same as before: the $b$-cycle integrations correspond to
integrals over the fermi sea, whereas the $a$-cycle integrations correspond to
integrals ``under the hill''. (These last ones should really be viewed as
regularized integrals, avoiding the singularity at $x=0$.)

The periods of $\gO$ can be calculated by exactly the same method as we used
before, and one finds
\bea
 \int_A \Omega & = & 2 \pi^2 \mu \ret
 \int_B \Omega & = & i \pi \left( \gL^2 - \mu_c \log (\mu_c / \gL^2) - \mubar_c
 \log (\mubar_c / \gL^2) \right)
 \label{eq:periodsX}
\eea
where we introduced the complexified $\mu_c = \mu + i \sqrt{M}$. Note that
the result does not depend on which of the possible $A$- or $B$-cycles we
choose. In particular, one might construct a more physical $a$-cycle along the
real axis in the $x$-plane (but avoiding the singularity by an imaginary
$\epsilon$) by adding two $a$-cycles; the resulting period will simply be $4
\pi^2 \mu$.

When we divide out $\bZ_2^A$ mapping $(p,x) \mapsto (-p,-x)$, nothing
happens to the fibers, and the base space will be divided out by the $\bZ_2^A$,
which as we saw acts within the leafs $\Ltilde_c$, turning them into the leafs
$L'_c$ of figure \ref{fig:mediumleaf}. Therefore, each cycle in X/2 corresponds
to the sum of two ``mirrored'' cycles in $X$, and the periods are simply twice
the periods in (\ref{eq:periodsX}).

\begin{figure}[ht]
 \begin{center}
  \includegraphics[height=5cm]{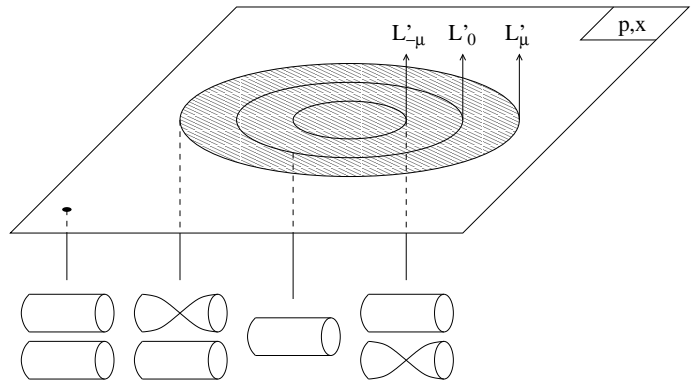}
 \end{center}
 \caption{The fibration giving the manifold $X'$. At the leafs $L'_{0, \pm \mu}$,
 the fiber degenerates. The shaded region represents a filled $a$-cycle of
 $L_\mu$. There are extra dimensions in this picture which result in the fact
 that we can either let this filled up cycle pass ``above'' $L'_0$ or ``under''
 $L'_0$, or, as we have drawn here, through it. In the latter case, in the
 interior of $L'_0$, we have the choice of two branches or the fiber.} 
 \label{fig:fibrationXp}
\end{figure}

Now, let us draw the ``double cover'' $X'$ of $X/2$:
\be
 X': \qquad \left( uv + \mu = \pm \left( \half p^2 - \half x^2 + \frac{M}{2
 x^2} \right) \right) / \bZ_2^A,
\ee
Note that the $\bZ_2$-action $(p,x) \mapsto (ip, ix)$ maps the leaf $L'_c$
to $L'_{-c}$. Over a general leaf $L'_c$, the fiber consists of the
two cylinders $uv = \pm c - \mu$. There are three special leafs: at
$L'_{\pm \mu}$ one of the two cylinders in the fiber becomes a pinched
cylinder, and at $L'_0$, the fiber becomes a single cylinder. This is
drawn in figure \ref{fig:fibrationXp}.

\begin{figure}[ht]
 \begin{center}
  \includegraphics[height=3cm]{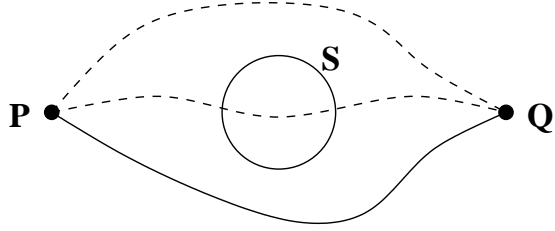}
 \end{center}
 \caption{The lower-dimensional analogue of the situation in figure
 \ref{fig:fibrationXp}. Even though the circle $S$ is one-dimensional, a
 one-dimensional curve from $P$ to $Q$ does not necessarily intersect it.
 However, when it does not, we have to choose on which side the curve passes
 $S$.} 
 \label{fig:curvecircle}
\end{figure}

Finding the $A$- and $B$-cycles in this geometry that correspond to the cycles
we identified in $X'$ requires some care. Let us focus on the $A$-cycles. Recall
that to construct the $A$-cycles, we have to take the $a$-cycles on $L'_{\pm
\mu}$, and view them as boundaries of disks in the full $(p,x)$-space. We now
have to ask ourselves whether this boundary intersects $L'_0$ or not. In
general, in a four-dimensional space, two infinite two-dimensional surfaces will
always intersect, but here, one of the directions on $L'_0$ is compact, so this
is not necessarily the case. A lower-dimensional analogue is drawn in figure
\ref{fig:curvecircle}: the curve from $P$ to $Q$ does not have to intersect the
circle $S$. However, we have to choose on which side of $S$ the curve passes,
and the same is true in the case under consideration here. Now we claim that the
two ways to ``fill up'' the cycle $a$ actually will lead to different
$A$-cycles. The easiest way to see this is to take the limiting situation where
the disk with boundary $a$ actually cuts through the cylinder $L'_0$, see
figure \ref{fig:fibrationXp}. On the inside of the circle on which the surfaces
intersect, there will then be a choice of two fibers, and these cannot be
continuously deformed into each other\footnote{One might wonder if a similar
two-fold degeneracy shows up depending on which side we use to pass the other
$L'_{\mp \mu}$, but since the fiber here still consists of two disconnected
components, one can continuously deform these two cycles into each other.}.

To confirm the above picture, one can actually carry out the cycle integrals
using the surface described above, and one then encounters a square root branch
cut where one has to choose a sign, corresponding to the choice of a fiber.
Instead of working this out in detail, we will calculate the periods (on
$X'/2$, which has the same behavior) in a much more elegant way in section
\ref{sec:0ACY}.

\begin{figure}[ht]
 \begin{center}
  \includegraphics[height=5cm]{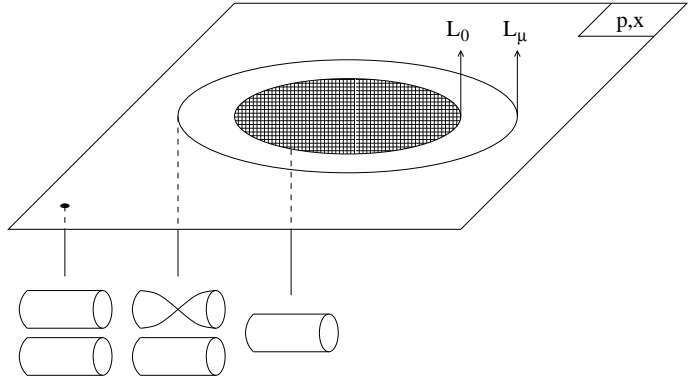}
 \end{center}
 \caption{The fibration giving the manifold $X'/2$. The shaded region is not
 part of the base space. Nevertheless, the fact that $L_0$ appears to be a
 boundary is an artifact of the picture; in reality, the space is perfectly
 regular there.} 
 \label{fig:fibrationXp2}
\end{figure}

Summarizing the story so far, we find that each $A$-cycle in $X/2$ corresponds
to four $A$-cycles in $X'$: two originating from $L'_{\mu}$ and two originating
from $L'_{-\mu}$. Finally, we have to project down $X'$ to $X'/2$, by using the
$\bZ_2^C$-action (\ref{eq:Z2C}). Since this action maps $L'_c$ to $L'_{-c}$,
this amounts to keeping half of the leafs, which now become of type $L_c$ as
explained above. The resulting picture is sketched in figure
\ref{fig:fibrationXp2}. Recall that the generic leafs are now simply cylinders,
which have one compact $a$- and one non-compact $b$-cycle. Again, we can
construct $A$-cycles by starting from the singe $a$-cycle on $L_\mu$. Note that
even though $L_0$ sits on the boundary of our picture, in reality the manifold
does not have a boundary, and we still have to worry about on which side of
$L_0$ our disk passes. Arguing like above, we therefore find two $A$-cycles
$A_\pm$, where the $\pm$ denotes ``over'' or ``under''. Similarly, we find two
B-cycles.

So the total picture is now as follows: each of the two cycles $A_\pm$ in $X'/2$
originates from two $A$-cycles in $X'$. However, {\em four} $A$-cycles of $X'$
project down to a single $A$-cycle in $X/2$. Therefore, an $A$-cycle in $X/2$
corresponds to the sum of the cycles $A_\pm$ in $X'/2$. We also saw that the
$A$-cycles in $X/2$ came from two $A$-cycles in $X$, so finally, we have that
the sum of two ``mirrored'' $A$-cycles in $X$ corresponds to the sum of $A_\pm$
in $X'/2$. However, because of the steps we had to take in between, we cannot
relate the {\em single} $A$-cycles on both sides. Comparing (\ref{eq:periodsX})
to (\ref{eq:Aperiods}) and (\ref{eq:Bperiods}), one can check that this picture
is correct. (The overall factor of two comes from a difference in normalization
of the natural (3,0)-forms on the two manifolds. The extra factor of two in the
relation of the $A$-cycles comes from the fact that one should really take two
$A$-cycles of $X$ to get a ``physical'' one.)

\end{document}